\documentclass[a4paper,11pt]{article}
\pdfoutput=1 
\usepackage{jcappub}
\usepackage{graphicx}
\usepackage{epsfig}

\title{IceCube Constraints on Fast-Spinning Pulsars as High-Energy Neutrino Sources}
\author[a]{Ke Fang}
\author[b]{, Kumiko Kotera}
\author[c]{, Kohta Murase}
\author[d]{and Angela V. Olinto}
\affiliation[a]{Department of Astronomy, University of Maryland, College Park, MD, 20742, USA}
\affiliation[b]{Institut d'Astrophysique de Paris, UMR 7095 - CNRS, Universit\'e Pierre $\&$ Marie Curie, 98 bis boulevard Arago, 75014, Paris, France}
\affiliation[c]{Department of Physics; Department of Astronomy \& Astrophysics; Center for Particle and Gravitational Astrophysics, The Pennsylvania State University, PA 16802, USA}
\affiliation[d]{Department of Astronomy \& Astrophysics, Kavli Institute for Cosmological Physics, The
  University of Chicago, Chicago, IL 60637, USA.}

\abstract{Relativistic winds of fast-spinning pulsars have been proposed as a potential site for cosmic-ray acceleration from very high energies (VHE) to ultrahigh energies (UHE). We re-examine conditions for high-energy neutrino production, considering the interaction of accelerated particles with baryons of the expanding supernova ejecta and the radiation fields in the wind nebula. We make use of the current IceCube sensitivity in diffusive high-energy neutrino background, in order to constrain the parameter space of the most extreme neutron stars as sources of VHE and UHE cosmic rays. We demonstrate that the current non-observation of $10^{18}$ eV neutrinos put stringent constraints on the pulsar scenario. For a given model, birthrates, ejecta mass and acceleration efficiency of the magnetar sources can be constrained. When we assume a proton cosmic ray composition and spherical supernovae ejecta, we find that the IceCube limits almost exclude their significant contribution to the observed UHE cosmic-ray flux. Furthermore, we consider scenarios where a fraction of cosmic rays can escape from jet-like structures piercing the ejecta, without significant interactions. Such scenarios would enable the production of UHE cosmic rays and help remove the tension between their EeV neutrino production and the observational data.}


\begin{document}

\maketitle 

\section{Introduction}

Galactic accelerators are likely responsible for the dominant component of cosmic rays observed on Earth (below $10^{15}\,$eV), given their containment by the Galactic magnetic field (e.g.,~\cite{Cesarsky80,Hillas84,Strong07}). The recent gamma-ray observations support that the main sources of these Galactic cosmic rays are supernova remnants \cite{Tavani10,Giuliani11,Ackermann13}. A transition is expected to occur at higher energies (around $10^{17-18}\,$eV), and cosmic rays should originate in -- yet unidentified -- extragalactic sources. The known Galactic objects do not possess the required energetics to produce cosmic rays above $10^{18}\,$eV. Besides, the presence of a source in the Galaxy contributing at these energies would induce a signature in the large-scale anisotropy of the arrival directions of cosmic rays, that is not observed. Measurements with the Auger Observatory and the Telescope Array are already constraining the Galactic-extragalactic transition energy and models of the Galactic magnetic field \citep{Giacinti11,Abreu12,Abu-Zayyad13}. 

At the highest energies, the possible candidate sources have been progressively narrowed down to a handful of objects over the last decades, but the major culprit has not been yet identified. Among the most promising sources, active galactic nuclei (AGN) with their black holes, jets, hotspots, and flares, as well as gamma-ray bursts (GRBs), including low-luminosity GRBs associated with trans-relativistic supernovae, are heavily plebiscited (see e.g., review~\cite{KO11} and references therein). A contender that was introduced early on by Refs.~\cite{Venkatesan97,Blasi00,Arons03} and has been resuscitated more recently by Refs.~\cite{Murase09,K11, FKO12,FKO13} are magnetized and fast-spinning neutron stars. These objects combine many advantages: their rotation speed endows them with a large energy reservoir ($E_{\rm rot}\sim 2\times10^{52}\,{\rm erg}\,I_{45}P_{-3}^{-2}$, with $I$ the star inertial momentum and $P$ its spin period\footnote{Here and in what follows, quantities are labelled $Q_x\equiv Q/10^x$ in cgs units unless specified otherwise.}, for an isolated new-born pulsar spinning close to the disruption limit), and their population density ($\dot{n}_{\rm s}\sim 10^{-4}\,{\rm Mpc}^{-3}\,{\rm yr}^{-1}$  \cite{Lorimer08}) is high enough to allow a comfortable total energy budget. The energy injected into UHECRs is of order ${\cal \dot{ E}}_{\rm UHECR} \sim 0.5\times 10^{44}\,{\rm erg}\,{\rm Mpc}^{-3}\,{\rm yr}^{-1}$~\cite{Murase:2008sa,Katz09}, which implies that a fraction of order $10^{-4}$ of the neutron star population is required to achieve the ultrahigh energy cosmic ray (UHECR) flux level. 

Within the zoo of neutron stars, those with extremely strong surface dipole fields of $B\sim 10^{15}\,$G, which are often called {\it magnetars} (see \cite{Woods06,Harding06,Mereghetti08} for reviews), have attracted particular attention because of their energetics \cite{Arons03,Murase09,K11}. This subpopulation, the existence of which was predicted in the 90s \cite{Duncan92}, is accepted as the explanation to the observed Soft Gamma Repeaters and Anomalous X-ray Pulsars \cite{Kouveliotou98,Kouveliotou99,Baring01}. 

Neutron stars are believed to be the byproducts of supernovae explosions. It is thus likely that, at the early stage of the neutron star life, when it is able to supply enough energy to accelerate particles to ultrahigh energies, it is surrounded by the radiatively and baryonically dense supernova ejecta. Assuming that particles can be successfully accelerated within the neutron star wind or nebular region, they will have to cross this interface, as well as the dense supernova ejecta, on their way to the interstellar medium. In Ref.~\cite{FKO12}, we demonstrated that for magnetars, the energy losses experienced by particles during their flight in the supernova ejecta did not allow their escape at ultrahigh energies, unless the ejecta mass was considerably lower than for standard core-collapse supernovae, or if a mechanism such as a jet was invoked to pierce the envelope. Indeed, magnetars spin down faster than more mildly magnetized stars, and the highest energy particles are produced when the surrounding ejecta is still too opaque to let them escape. 

Magnetars are thus not necessarily favored to produce cosmic rays at the highest energies. They could be contributing however at energies below $\sim 10^{18-19}\,$eV. In this energy range, the composition measured by the current experiments indicate that protons are dominant~\cite{AugerIcrc11,TAicrc11}. 

The interactions of cosmic rays within the nebula or supernova ejecta regions should lead to the generation of secondary particles, including high energy neutrinos, which has been suggested as a very powerful test of newborn pulsar scenarios for UHECRs \cite{Murase09}.
The guaranteed level of neutrinos expected for a pulsar population fitting the Auger data (both spectrum and composition) was calculated in Ref.~\cite{FKMO14_letter}. We demonstrated there that this flux would be observable by IceCube within the next ten years. 

In this work, we make use of the current IceCube sensitivity in neutrinos, in order to constrain the parameter space of the most extreme neutron stars as extragalactic sources of cosmic rays, focussing on the particle energy range above $10^{17}\,$eV. As discussed in this study, this range corresponds indeed to the peak of the produced neutrino flux. We demonstrate that the current non-observation of neutrinos in the EeV range radically shrinks the range of parameters allowed for magnetars that would efficiently produce very high-energy cosmic rays, and almost exclude any contribution from these sources. For higher energies, as mentioned above, either a low mass supernova ejecta or a jet-like structure is needed to let particles escape without too much damage. We show that even within these scenarios, the constraints imposed by the level of neutrinos at EeV energies still partially constrain magnetars as sources of UHECRs.

We give in Section~\ref{section:scenario} an overview of the UHECR production issues (ion injection, acceleration and escape) related to neutron stars and their surrounding nebula and supernova ejecta. We estimate the corresponding neutrino fluxes in Section~\ref{section:neutrinos}. In Section~\ref{section:param}, we present our parameter scan over the neutron star pulsation and dipole field, the source birth rate, and the particle acceleration efficiency. In Section~\ref{sec:jet} we consider scenarios with the presence of  jet-like structures.  We discuss our results and conclude in Section~\ref{section:conclusion}.

\section{UHECRs and High-energy neutrinos from young neutron stars}\label{section:scenario}

\subsection{Particle injection and acceleration}

Ions (from light to heavy nuclei) can be stripped off the neutron star surface by a combination of strong electric fields and bombardment of particles \cite{Ruderman75,Arons79}. 
The acceleration mechanism of these extracted particles up to ultrahigh energies in neutron star  environments is an unclear point of this source scenario. Our poor knowledge of the neutron star magnetospheres, winds, nebulae and termination shocks (at the interface between the wind and the surrounding supernova ejecta) is central to the difficulties encountered in building a detailed acceleration model, consistent with the observations and the leptonic emission counterparts. The features of the radiation due to pairs are themselves challenging to explain, and despite an increasing experimental and theoretical effort been put to understand the working of neutron star outflows and nebular emissions, the community is still struggling to solve fundamental problems (see e.g., reviews by Ref.~\cite{Arons02,Kirk09}), such as how and where pairs are been accelerated, or the related problem of electromagnetic to kinetic energy transfer within the wind (the so-called $\sigma$-problem). 

One certainty however is that neutron stars spin down, and subsequently, their rotational energy is channeled via their winds towards the outer medium. Following Ref.~\cite{LKP13}, one can calculate that particles of mass number $A$ in the wind can reach a maximum energy at neutron-star birth
\begin{equation}\label{eq:E0}
E_0 \sim 2.7\times10^{20}\,{\rm eV} A\,\eta\,\kappa_4^{-1}P_{{\rm i},-3}^{-2}B_{15}R_{\star,6}^3 \ ,
\end{equation}
where $B$ is the dipole magnetic field strength of the star, $R_\star$ its radius and $P_{\rm i}$ the initial spin period. The value of $\kappa$, the pair multiplicity, can range between $10-10^8$ in theory (a highly debated quantity, see e.g., \cite{Kirk09}).  In this work we take $\kappa\sim10^4$, which means that $\sim$10\% of the pulsar power goes into ions, if  particles receive the full potential of the pulsar. This energy conversion efficiency is consistent with the prediction  by recent simulations of pulsar magnetospheres \citep{2041-8205-795-1-L22}. 
 This estimate assumes that the electromagnetic luminosity of the pulsar wind $L_{\rm p}\sim\,6.6\times10^{48} P_{-3}^{-4}B_{15}^2R_{\star,6}^6\,$erg/s is converted into kinetic luminosity $\dot{N}mc^2$ with efficiency $\eta\le 1$: $E_0 = \eta L_{\rm p}/\dot{N}mc^2$. 
The particle rest mass power can be written as a function of the Goldreich-Julian rate $\dot N_{\rm GJ}$ \cite{Goldreich69}: $\dot N mc^2 \,\equiv\,\dot N_{\rm GJ} ( 2\kappa\,m_e + A m_{\rm p}/Z) c^2$ \cite{LKP13}. 
 
In equation~\ref{eq:E0} we assume that the flux of particles being heated by the pulsar follows   the Goldreich-Julian rate $\dot{N}_{\rm GJ}$, as we consider the scenario that ions only get accelerated in the pulsar wind. If the magnetic energy is dissipated in a volumetric way throughout the nebula by e.g., reconnection  \citep{2013MNRAS.431L..48P},  pairs could be created in abundance in the nebula, in particular for rapidly rotating millisecond pulsars \citep{2014MNRAS.437..703M}. However, it is unclear if the magnetic energy in the nebula could be dissipated efficiently into the pairs. 

We have also assumed  a proton composition of cosmic rays.  Some mechanisms (such as extraction of ions by strong electric fields \citep{Ruderman75, Arons79}, mixing of the stellar material via Kelvin-Helmholtz instabilities or oblique shocks \citep{Murase08,Wang08}, nucleosynthesis at the proto-pulsar phase \cite{2011MNRAS.415.2495M,Horiuchi12}) could lead to heavy nuclei injection, as we have discussed in our earlier work \cite{FKO12}.

Note that acceleration to high energies can only happen in the first stages of the life of the neutron star, typically within the spin-down timescale 
\begin{equation}
t_{\rm EM} = \frac{9Ic^3P^2}{8\pi^2B^2R^6} \sim 3.1\times 10^{3}\,{\rm  s}\,I_{45}B_{15}^{-2}R_{\star,6}^{-6}P_{{\rm i},-3}^2\ . 
\end{equation}
If gravitational wave losses are substantial, the star spins down over a timescale 
\begin{equation}
t_{\rm GW}=\frac{5c^5P^4}{2^{10}\pi^4GI\epsilon^2}\sim 1.5\times 10^{6}\,{\rm s}\, P_{\rm i,-3}^4I_{45}^{-1}\epsilon_{-4}^{2}\ ,
\end{equation} 
with $\epsilon$ the ellipticity created by the interior magnetic fields of the star,  if the magnetic distortion axis and the rotation axis of the star are not aligned \cite{Usov92,Bonazzola96,Ostriker69}. Typically $\epsilon =\beta R_\star^8B^2/(4GI^2)\sim  4\times10^{-4}\beta_2R_{\star,6}^8B_{15}^2I_{45}^{-2}$, where $\beta$ is the magnetic distortion factor introduced by Ref.~\cite{Bonazzola96}, which measures the efficiency of the interior magnetic field in distorting the star. This factor depends on the equation of state of the star interior and on its magnetic field geometry. Ref.~\cite{Bonazzola96} finds that the value of $\beta$ can range between $1-10$ for perfectly conducting interiors (normal matter), $10-100$ for type I superconductors and can reach $\gtrsim 100$ for type II superconductors (for a detailed study on the connection between magnetars as sources of UHECRs and gravitational waves, see Ref.~\cite{K11}). We will note 
\begin{equation}
t_{\rm sd}\equiv (t_{\rm EM}^{-1}+t_{\rm GW}^{-1})^{-1}\ .
\end{equation} 

In the following, we will place ourselves in a regime where the conversion efficiency of the wind electromagnetic into kinetic luminosity is high enough to achieve ultrahigh energies. Our results remain valid if we relax this assumption, at the cost of a softer injection spectrum that would be produced by the stochastic acceleration to reach the highest energies.\footnote{ In the case when the conversion is not fully efficient, stochastic types of acceleration could take place at the shock to further push the maximum acceleration energy to the confinement limit  $\gamma_{\rm conf} = {Ze B_{\rm PWN} R_{\rm PWN}}/{(A m_{p} c^2)}$ (where $B_{\rm PWN}$ and $R_{\rm PWN}$ represent respectively the pulsar wind nebula magnetic field and radius), which can reach values $>10^{11}$ over the spin-down timescale for neutron stars with parameters $B\gtrsim 10^{12}\,$G and initial rotation period $P_{\rm i}\sim 1\,$ms \cite{LKP13}.  See however our discussion on acceleration limits due to synchrotron cooling a few paragraphs below.
}
Taking into account the neutron-star spin down (assuming a breaking index of 3, corresponding to a spin-down luminosity $L_{\rm sd}=L_p\,(1+t/t_{\rm sd})^{-1}$), we thus consider that cosmic rays are accelerated at a given time $t_{3.5}=t / 10^{3.5}\,\rm s$ at energy~\cite{Blasi00,Arons03}
\begin{eqnarray}\label{eqn:Et}
E_{\rm CR} (t) &=& E_0\,\left(1+t/t_{\rm sd}\right)^{-1} \nonumber \\
&\sim& 1.3\times10^{20}\,{\rm eV} \,A\,\eta\,\kappa_4I_{45}B_{15}^{-1}R_{\star,6}^{-3} \,{t_{3.5}}^{-1}\,\ .
\end{eqnarray}
Channeling the Goldreich-Julian charge density into particles and taking into account the neutron-star spin down rate, one can write the cosmic-ray injection flux \cite{Blasi00,Arons03}
\begin{equation}\label{eq:spectrum_arons}
\frac{{\rm d} N_{\rm CR}}{{\rm d} E}(t) = \frac{9}{4}\frac{c^2I}{ZeBR_\star^3} E_{\rm CR}(t)^{-1}\left[ 1+\frac{E_{\rm CR}(t)}{E_{\rm GW}}\right]^{-1}\, ,
\end{equation}
where $E_{\rm GW}$ is the critical gravitational energy at which gravitational wave and electromagnetic losses are equal. The cosmic-ray luminosity then reads
\begin{equation}
L_{\rm CR}(t) \equiv E_{\rm CR}^2\frac{{\rm d}^2 N_{\rm CR}}{{\rm d} E\,{\rm d}t}(t) = \frac{9}{4}\frac{c^2I}{ZeBR_\star^3} E_{\rm CR}(t)(t+t_{\rm sd})^{-1}\, .
\end{equation}

\subsection{Radiation backgrounds in the nebula region and in the supernova ejecta}\label{sec:radiation}

Ultrahigh energy ions can experience photo-pion production or photo-disintegration in the radiation fields surrounding the neutron star: in the nebular region at the interface between the pulsar wind and the supernova shell (a non-thermal component), and in the supernova ejecta, where most of the incident non-thermal radiation is thermalized over short timescales~\cite{KPO13}. 

For neutrino production, the effects of this non-thermal radiation background in the nebula can be neglected as long as the confinement time of UHECRs in this region $t_{\rm conf, neb}$ is shorter than the photo-hadronic interaction timescale $t_{A\gamma,{\rm neb}}$. The most important contribution to pion-production will come from photons produced by synchrotron emission. The synchrotron photon density for nebulae of fast-spinning neutron stars, $n_\gamma$, can be computed following  Ref.~\cite{LKP13}, and the corresponding interaction timescale reads $t_{A\gamma,{\rm neb}} = (\sigma_{A\gamma}n_{\gamma}c)^{-1}$, where $\sigma_{A\gamma}$ is the photo-hadronic interaction cross-section. The magnetic confinement timescale in the nebula can be expressed $t_{\rm conf, neb} = R_{\rm neb}^2/(r_L c)$, assuming a Bohm diffusion regime in the nebula magnetic field of strength $B_{\rm neb}$.

\begin{figure}[tbp]
\centering
\includegraphics[width=0.7\textwidth]{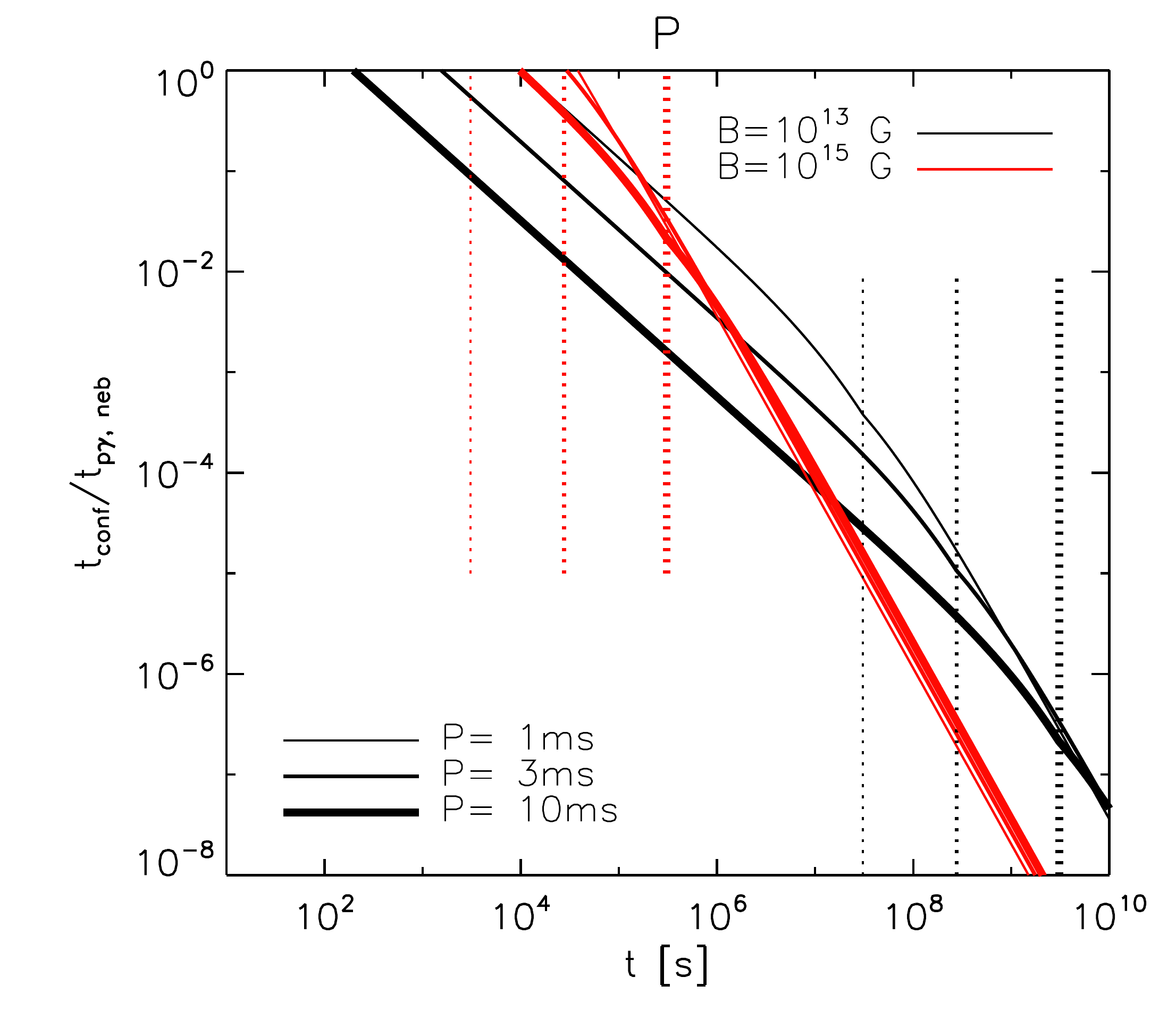} 
\caption{Time evolution of the ratio of the confinement timescale, $t_{\rm conf}$, to the pion production timescale, $t_{p\gamma}$, in the non-thermal radiation field of the nebula region, for a proton at Lorentz factor $10^{9}$, for a neutron star with initial rotation period $P_{-3}=1, 3,10$ (increasing thickness), dipole magnetic field $B_{\rm \star, 13}=1,100$ (black and red), leptonic multiplicity $\kappa_{4}=1$ and $\eta_{\rm B}=0.1$. The vertical dotted line indicates the spin-down timescale $t_{\rm sd}$  corresponding to each rotation period (increasing thickness).  \label{fig:nebula}}
\end{figure}

These timescales depend on the size and on the average strength of the magnetic field in the expanding nebula region. These quantities can be calculated applying the estimates of Ref.~\cite{Chevalier05} (see also \cite{KPO13, LKP13}). Figure~\ref{fig:nebula} illustrates that the ratio $t_{\rm conf, neb}/t_{A\gamma,{\rm neb}} \ll 1$ for fast-spinning pulsars and magnetars. This is in particular valid at times close to $t_{\rm sd}$ (vertical dotted lines) when most of the neutron star luminosity is provided. In the calculations, we have chosen $\eta_B=0.1$, a parameter corresponding to the magnetic fraction of the energy injected into the nebula (and that is actually contained in $\eta$ in Eq.~\ref{eq:E0}). This value is rather conservative as $\eta_B \ll 1$ from the observations of pulsar wind nebulae (e.g., \cite{Rees74,Kennel84,Kennel84b,Atoyan96,DelZanna04,Komissarov04}). These figures were computed for a particle Lorentz factor of $10^9$, but lower energies would lead to even milder effects of the radiation field.

As was pointed out in Ref.~\cite{KPO13}, it is thus possible that the UHECRs cross the nebula region without undergoing efficient photo-hadronic interactions. This remains valid except for strongly magnetized neutron stars. Note that protons would also cool via synchrotron in the strong magnetic field of the nebula, then particles would not reach very high energies for strongly magnetized magnetars \cite{LKP13}.  
However, UHECRs could interact with the radiation field of the supernova and produce ultrahigh energy neutrons that are not affected by synchrotron cooling. Besides, the conversion of the wind electromagnetic energy to kinetic energy is often assumed to be   efficient because of the observation of the Crab Nebula \cite{Kirk09}, but this is not certain at the early stages and in the types of objects we are considering. This point is further supported by the fact that the MHD theory is unable to reproduce an efficient conversion (the so-called $\sigma$-problem) \cite{Kirk09}.  This would imply that the magnetic energy density of turbulent fields could be small enough, where synchrotron cooling can be avoided. Another possibility is that in a typical Type Ibc supernova, the ejecta mass could be as small as $\sim 2 M_\odot$ so that the ejecta velocity is larger, creating a larger nebula size with milder magnetic fields. Therefore cosmic ray particles may avoid significant synchrotron losses and reach the highest energies.

Particles then enter the supernova ejecta. At times shorter or of order the spin-down time $t_{\rm sd}$ that we consider here, the supernova ejecta is dense enough to provide a non-negligible interaction medium for the accelerated cosmic rays. In particular, the incident non-thermal radiation field from the nebula is nearly immediately thermalized and provide a thermal radiation background for photo-hadronic interactions.

The ejecta of a standard Type II core-collapse supernova can be modeled as a sphere expanding with a velocity \cite{KPO13}
\begin{equation}\label{eq:vej}
\beta_{\rm ej} = \left(\frac{2E_{\rm SN}}{M_{\rm ej}\,c^2}\right)^{1/2}\left(1+\frac{E_{\rm rot}}{E_{\rm SN}}\right)^{1/2}\sim 4.8\times 10^{-2}\,I_{45}^{1/2}P_{\rm i,-3}^{-1}M_{\rm ej,1}^{-1/2}\ ,
\end{equation}
where the explosion energy of the supernova ejecta $E_{\rm SN}=10^{51}\,$erg and the pulsar wind energy $E_{\rm rot}=2\pi^2 I/P_{\rm i}^2$. The size of the ejecta can be written as $R_{\rm ej}(t)=\beta_{\rm ej} c t$.  Notice that for clarity, from here on we do not show the dependence of the numerical estimates on the inertia and the star radius (set to $I_{45}$ and $R_{\star,6}$ respectively), as these quantities are well-fixed by neutron star physics.

The thermal photon energy density in the supernova ejecta $U_{\rm th}$ reads 
\begin{equation}
U_{\rm th} = \frac{3\,E_{\rm th}}{4\pi R_{\rm ej}^3}
\end{equation}
where  $E_{\rm th} = E_{\rm SN, th} + \eta_{\rm th}\,L_{\rm sd}\,t$, with $E_{\rm SN, th}\sim 10^{49}\,\rm erg$ being the thermal energy from the heating by supernova shocks and unstable isotopes such as $^{56}$Ni, 
 and   $\eta_{\rm th}$ is the fraction of the rotational energy  converted into thermal photons   in the ejecta \citep{Murase:2007yt}, which we assume to be $\eta_{\rm th}=1$. 
For comparison,  thermalization of photons in the nebula usually happens when the optical depth of Thomson scattering gets around $\sim10-100$, corresponding to a $1-10$ \% conversion from the rotational energy to thermal photons.  
 We note a more detailed modeling of the thermal photon energy density in \cite{2015arXiv151205368L}.  
The corresponding ejecta temperature can be expressed $T = ({U_{\rm th}}/{a})^{1/4}$, where $a$ is the radiation constant. 
This thermal component peaks at energy $\epsilon_{\gamma,{\rm th}}=kT$.
The thermal radiation background leads to a cooling time by photo-pion interaction for a proton 
\begin{eqnarray}\label{eqn:t_pgamma}
t_{p\gamma,{\rm th}} &=&  \left(c\,\xi_{p\gamma}\sigma_{p\gamma}\frac{U_{\rm th}}{\epsilon_{\gamma, {\rm th}}}\right)^{-1}\nonumber \\
&\sim& 8\times 10^{-6}\,{\rm s}\, \eta_{\rm th,0}^{-3/4}P_{\rm i,-3}^{-3/4}M_{\rm ej,1}^{-9/8}t_{3.5}^{9/4} 
\end{eqnarray}
where  $\sigma_{p\gamma}\sim 5\times 10^{-28}\,$cm$^{-2}$, and the elasticity of the $p\gamma$ interaction $\xi_{p\gamma}=0.2$ \citep{Murase09}. In this calculation, we have assumed that the particle energy lies always above photo-pion interaction threshold (and not in the resonance peak of the $p\gamma$ interaction cross-section).
For analytical estimations here and in the rest of this section, we have assumed parameters for which the pulsar rotational energy dominates the supernova intrinsic thermal energy.

 The opacity then reads
\begin{eqnarray}
\tau_{p\gamma,{\rm th}} &=&  \frac{R_{\rm ej}}{ct_{p\gamma}}\sim 1.9\times 10^{7}\, \eta_{\rm th,0}^{3/4} P_{\rm i,-3}^{-1/4}M_{\rm ej,1}^{5/8}t_{3.5}^{-5/4}  \ ,
\end{eqnarray}
and the time at which the medium becomes transparent to protons
\begin{eqnarray}
t_{p\gamma,{\rm th}}^* \equiv t(\tau_{p\gamma,{\rm th}}=1)\sim 2.1\times 10^{9}\,{\rm s}\,\eta_{\rm th,0}^{3/5} P_{\rm i,-3}^{-1/5}M_{\rm ej,1}^{1/2} \ .
\end{eqnarray}
From these estimates, one can see that contrarily to the non-thermal radiation field in the nebula, the thermal radiative background in the supernova ejecta should have a strong effect on the accelerated particles at early times. Only a considerably small value of $\eta_{\rm th}$ would minimize its effect. 
 
Cosmic ray energy at time $t_{p\gamma,{\rm th}}^*$ reads
\begin{equation}
E_{\rm CR}(t=t_{p\gamma,\rm th}^*) = 3.6\times10^{14}\,{\rm eV}\, A\eta\kappa_4 B_{15}^{-1}\eta_{\rm th, 1}^{-3/5}P_{i,-3}^{1/5}M_{\rm ej,1}^{-1/2} \ .
\end{equation}

We define
\begin{equation}
t_{p\gamma} \equiv \min(t_{p\gamma, {\rm th}},t_{p\gamma, {\rm neb}}) \ ,
\end{equation}
its corresponding opacity $\tau_{p\gamma}$, and the time at which the medium becomes transparent for protons $t_{p\gamma}^*\equiv t(\tau_{p\gamma}=1)$. From the above discussion, $t_{p\gamma}^*\sim t_{p\gamma, {\rm th}}^*$.

\subsection{Surrounding ejecta and escape (baryonic background)}

At early times, the supernova ejecta presents also a dense baryonic background that can lead to efficient hadronic interactions for UHECRs. The mean density of the sphere over the size $R_{\rm ej}(t)=\beta_{\rm ej} c t$ can be written \cite{Chevalier05}:
\begin{equation}\label{eq:rhoSN}
\rho_{\rm ej}(t)=\frac{3\,M_{\rm ej}}{4\pi \beta_{\rm ej}^3c^3t^3} \sim 5.1 \times 10^{-5}\,M_{\rm ej,1}^{5/2}P_{\rm i,-3}^{3} t_{3.5}^{-3}~{\rm g\,cm}^{-3}\,\ .
\end{equation}
Here $M_{\rm ej}=10\,M_{\rm ej, 1}\,M_\odot$ denotes the ejecta mass and $E_{\rm ej}=E_{\rm rot}+E_{\rm exp}\sim 2\times10^{52}\,{\rm erg}\,I_{45}P_{\rm i,-3}^{-2}$ the ejecta energy, expressed as the sum of the neutron star rotational energy and the supernova explosion energy ($E_{\rm exp}\sim 10^{51}\,$erg) \citep{Matzner99}. Equation~(\ref{eq:rhoSN}) provides a reasonable estimate of the evolution of the density of the ejecta surrounding the neutron star for various supernova scenarios, as is discussed in Ref.~\cite{FKO12}. Note that results are only mildly sensitive to the ejecta mass within a range $5-20\,M_\odot$. 

As discussed in Ref.~\cite{FKO12}, the composition of the supernova ejecta depends upon the type, progenitor mass, and the final interior mass of the supernova, and rotation-powered pulsars and magnetars have been invoked as the remnant of a wide variety of supernova types, such as Ib, Ic, or II. The composition of all these objects is dominated by Hydrogen or light elements. For simplicity, we will consider in the following that the ejecta is composed of pure Hydrogen. We have demonstrated in Ref.~\cite{FKO12} that different ejecta composition did not have a considerable impact as far as the production and escape of UHECRs was concerned. 

The hadronic interaction timescale can be expressed as
\begin{equation}\label{eqn:t_pp}
t_{pp} = m_{p}(c \,\rho_{\rm ej}\sigma_{pp}\xi_{pp})^{-1} \sim 1.7\times 10^{-5}\,{\rm s} \,M_{\rm ej,1}^{-5/2} P_{\rm i,-3}^{-3} t_{3.5}^{3}\ ,
\end{equation}
and the background optical depth for protons
\begin{equation}
\tau_{pp}=\frac{R_{\rm ej}}{ct_{pp}}\sim 8.9\times 10^{6}\,M_{\rm ej, 1}^{2}\, P_{\rm i,-3}^{2}\,t_{3.5}^{-2} \ ,
\end{equation}
with the proton-proton interaction cross-section $\sigma_{pp}\approx1\times 10^{-25}\,\rm cm^2$ at $10^{18}$ eV and $\xi_{\rm pp} \sim 0.5$. Note that, at a given time, this quantity does not depend on the magnetic field of the neutron star. The dependence is actually implicitly considered, as we will evaluate the optical depth at time $t_{\rm sd}$, which is shorter for faster spin-downs, i.e., for higher $B$. The corresponding time and energy at which cosmic rays can escape read
\begin{equation}
t_{pp}^*\equiv t(\tau_{pp}=1)\sim 9.4\times10^{6}\,{\rm s}\,M_{\rm ej, 1} P_{\rm i,-3}\ .
\end{equation}
\begin{equation}
E_{\rm CR}(t=t_{pp}^*)\sim 6.0\times10^{16}\,{\rm eV}\,\eta\kappa  P_{\rm i,-3}^{-1}B_{15}^{-1} M_{\rm ej,1}^{-1}\ .
\end{equation}
This equation shows that, as was demonstrated and discussed in Refs.~\cite{Blasi00,FKO12}, the ejecta is too dense to allow the escape of the highest energy protons through the supernova ejecta. For milder magnetic fields than for magnetars, the spin-down time is longer and the ejecta can become diluted enough to allow the escape of heavy nuclei at UHE. One might consider however that some mechanisms invoked  in the next section and in Section~\ref{sec:jet} can carve a path for cosmic rays to escape safely, for magnetars in particular which are extreme objects. \\

\subsection{Neutrino production and diffuse flux}\label{section:neutrinos}

\begin{figure}[t]
\centering
\epsfig{file=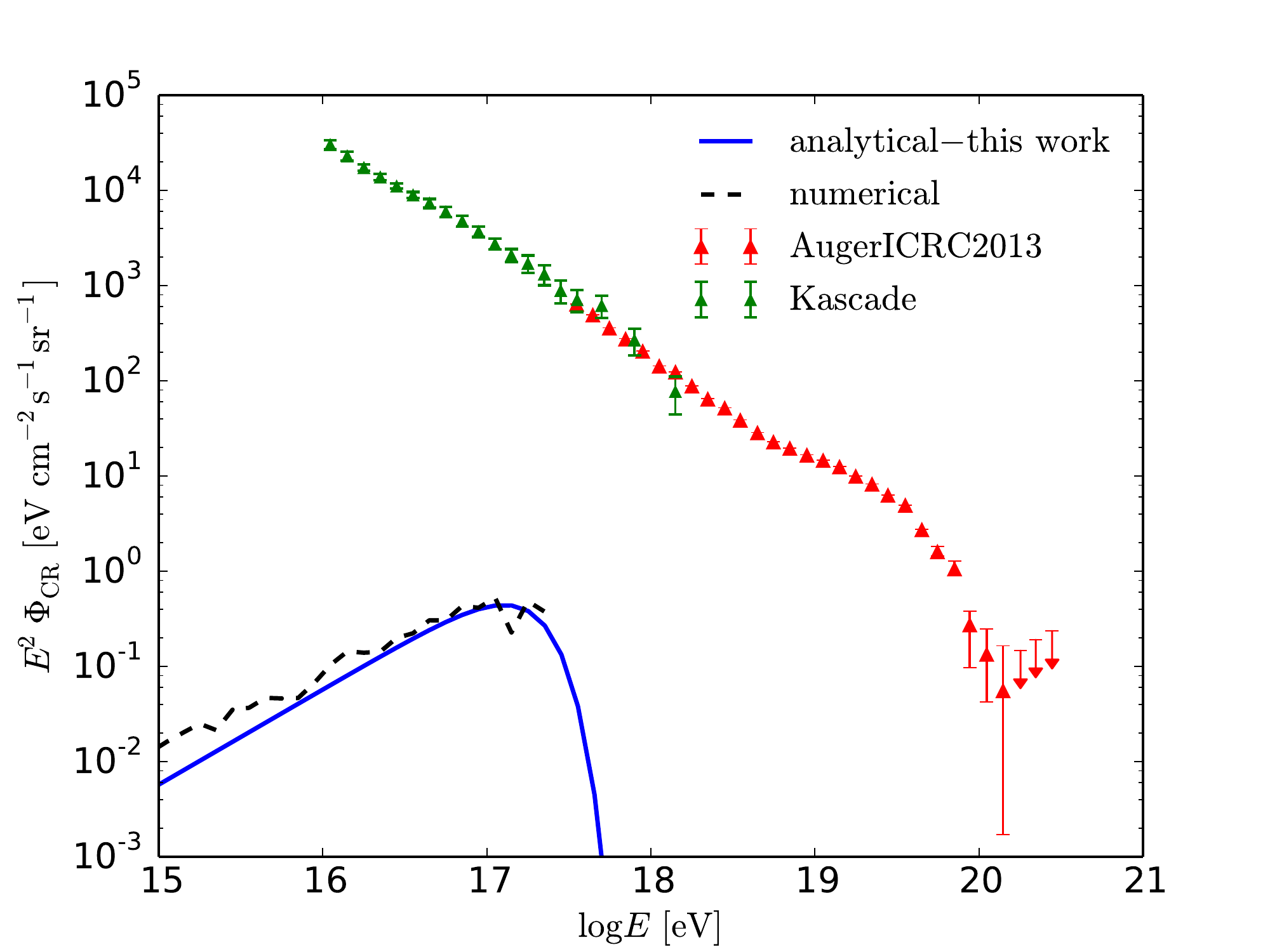,width=0.48\textwidth}
\epsfig{file=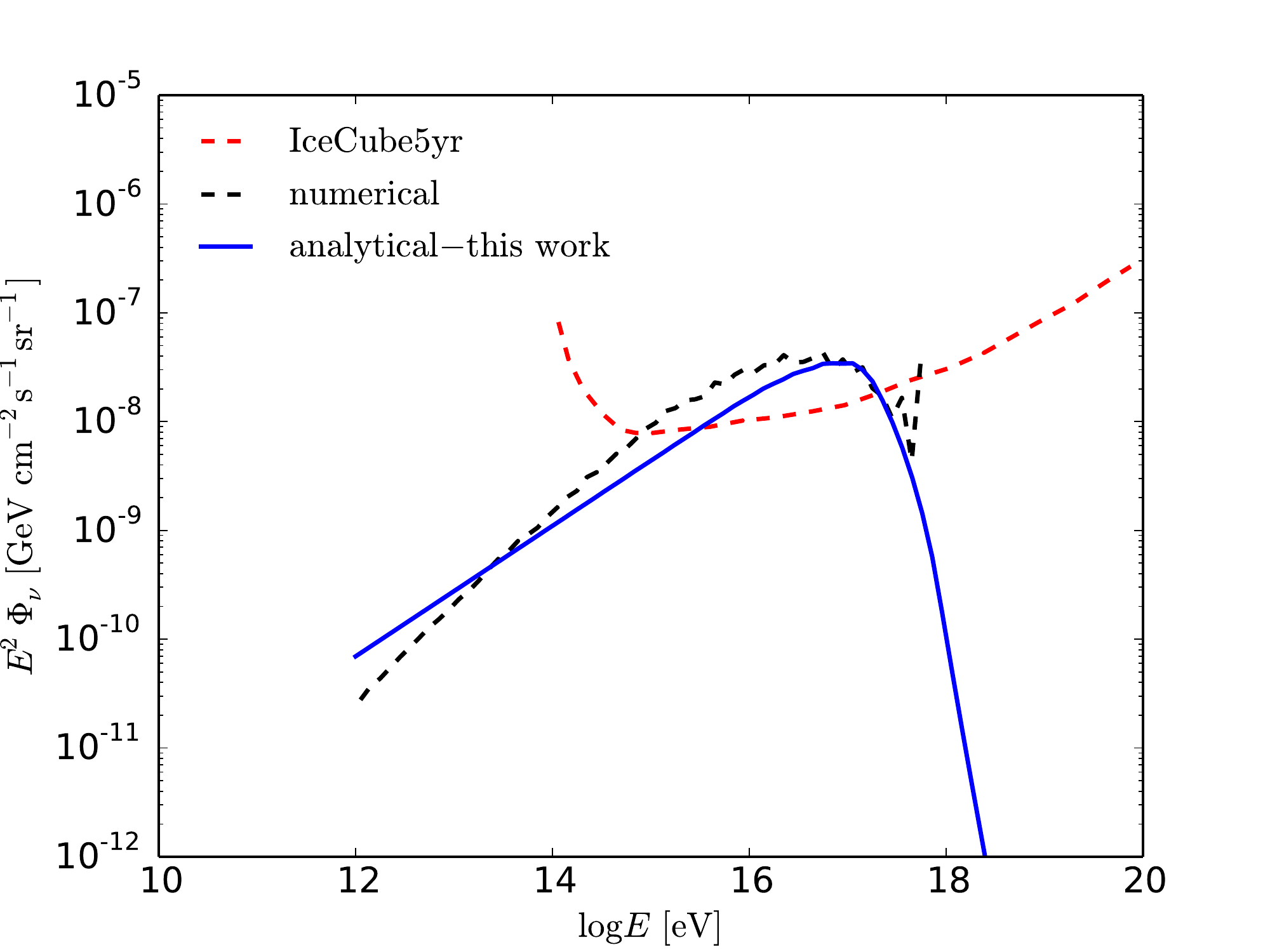,width=0.48\textwidth}
\caption{Spectra of cosmic rays (left) and neutrinos (right) from a magnetar with surface magnetic field $B=10^{15}$ G and initial spin period $P_i=1$ ms, comparing to measurements of the KASCADE \citep{Apel:2011mi, Apel:2013ura} and the Auger Observatory \citep{ThePierreAuger:2013eja}, as well as the IceCube 5-year sensitivity  \citep{Abbasi:2011zx, 2013ApJ...779..132A}. Only hadronuclear interaction is taken into account, and the acceleration efficiency is set to be $\eta=1$. In both plots,   analytical results from this work (blue) are comparable to numerical results from \cite{FKO12, F15} (note that in the right panel, the  black dashed line shows only neutrinos from primary cosmic rays). }
\label{fig:spec} 
\end{figure}

The cosmic-ray interactions on the hadronic and radiative backgrounds described in the previous sections will inevitably lead to the production of charged pions, and thus of neutrinos. The meson production efficiency reads 
\begin{equation}
f_{\rm mes}=\min{(\tau_{pp}  + \tau_{p\gamma} , 1)}
\end{equation}
We will assume that for each interaction, charged pions undertake a fraction of the parent cosmic-ray energy $f_\pi\equiv E_\pi/E_{\rm CR}\sim 0.2\,$, and each neutrino $f_\nu\equiv E_\nu/E_{\pi}\sim 0.25$. 

At early times when the ejecta is very dense, the secondary nuclei and pions continue to interact with the radiation and hadron background and produce higher order nuclei, neutrinos and pions \cite{Murase09}. Charged pions have a lifetime of $\tau_\pi=2.6\times10^{-8}\,$s in the lab frame. They interact with protons with cross section $\sigma_{\pi p} \sim 5 \times 10^{-26}\,\rm cm^2$ and elasticity $\xi_{\pi p}=0.5$  \cite{Murase09}, and with thermal photons with $\sigma_{\pi\gamma}\sim 10^{-28}\,\rm cm^2$ and $\xi_{\pi\gamma}\sim 0.5$, producing additional neutrinos and pions that undergo further $\pi p$ and $\pi\gamma$ interaction. Notice that  the  $\pi\gamma$ cross section was estimated by $\sigma_{\pi \gamma}\sim\sigma_{p\gamma}\,(\sigma_{\pi p}/\sigma_{pp})$. 
This cascade continues until $\min(t_{\pi\gamma},t_{\pi p})={\gamma_\pi\,\tau_\pi}$. Charged pions then stop interacting and decay into neutrinos via $\pi^\pm\rightarrow e^\pm+\nu_e(\bar{\nu}_e)+\bar{\nu}_\mu +\nu_\mu$. One can then define a neutrino flux suppression factor as 
\begin{equation}
f_{\rm sup}=\min\left[1,\left( \left(\frac{t_{\pi p}}{\gamma_\pi\,\tau_\pi}\right)^{-1} +  \left( \frac{t_{\pi \gamma}}{\gamma_\pi\,\tau_\pi}\right)^{-1}   \right)^{-1} \right] \sim 6.5\times 10^{-8}\,\eta^{-1}\kappa_4^{-1}P_{\rm i,-3}^{-3}B_{15}M_{\rm ej,1}^{-5/2}t_{3.5}^4 \ .
\end{equation}

Assuming that cosmic rays follow  $dN_{\rm CR}/dE_{\rm CR}\propto E_{\rm CR}^{-p}$ , then the spectrum of the neutrinos from the pion decay  can be written as \cite{Murase09}:
\begin{equation}
E_\nu^2\frac{{\rm d}N_\nu}{{\rm d}E_\nu} \propto \begin{cases} \left(E_\nu/E_{\nu}^{\rm had}\right)^{(2-p)} &\mbox{if } E_\nu<E_\nu^{\rm had} \\ 
 \left(E_\nu/E_{\nu}^{\rm had}\right)^{(1-p)} e^{-E_\nu/(f_\nu\,E_{\rm CR})} & \mbox{if }  E_\nu^{\rm had} < E_\nu<E_{\rm CR}/4 \end{cases}
\end{equation}
where $E_{\nu}^{\rm had}\approx 0.25 \,\left(t_{\pi p}(E_{\rm CR})/\tau_\pi\right)\,m_\pi c^2$ is the break energy for cosmic rays injected with energy $E_{\rm CR}$. The neutrino flux is normalized by 
\begin{equation}
\int E_\nu\frac{{\rm d}N_\nu}{{\rm d}E_\nu} dE_\nu =\int \frac{3}{8} E_{\rm CR}\, \frac{{\rm d}N_{\rm CR}}{{\rm d}E_{\rm CR}}\, f_{\rm sup}f_{\rm mes} dE_{\rm CR}\ .
\end{equation}

The total neutrino spectrum thus breaks for $f_{\rm sup}=1$ at time
\begin{equation}\label{eqn:t_b}
t_{\nu,\rm b}=3.0\times 10^5 \,{\rm s}\,\eta^{1/4}\kappa_4^{1/4}P_{\rm i,-3}^{3/4}B_{15}^{-1/4}M_{\rm ej,1}^{5/8}.
\end{equation}
Inserting $t_{\nu,\rm b}$ into equation~\ref{eqn:Et},  the break is found at energy
\begin{equation}
E_{\nu,\rm b}= 1.2\times 10^{17}\,{\rm eV}\,A\kappa_4^{3/4}\eta^{3/4} P_{\rm i,-3}^{-3/4}B_{15}^{-3/4}M_{\rm ej,1}^{-5/8} \ .
\end{equation}
For the two estimates above, we have assume that $\pi p$ interactions are dominant over $\pi\gamma$, as $t_{\rm \pi p} < t_{\rm \pi\gamma}$  for our chosen parameters $t_{3.5}$, $B_{15}$ and $P_{i,-3}$. 
Note that most of the neutrinos are produced by cosmic rays of  $10^{18-19}$ eVs.
The neutrino flux at the break energy from a population of UHECR sources with birth rate $\Re(D)$ at a given distance $D$, can be estimated as
%\begin{eqnarray}
%E_{\nu,\rm b}^2\,\Phi_{\nu,\rm b}&=&\frac{1}{4\pi}\,\Re(0)\,f_z D_{\rm H}  \int_0^{t_{\nu,b}} {L_{\nu}}\,{\rm d}t' \\
%&\approx& \frac{1}{4\pi}\,\frac{3}{8}\,\Re(0)\,f_z D_{\rm H}  \frac{dN_{\rm CR}}{dE}(0)\,\frac{t_{\nu,\rm b}}{t_{\rm sd}}\,E_{\rm CR}^2(t_{\nu,%\rm b}) \,f_{\rm mes}\,f_{\rm sup}
%\end{eqnarray}
\begin{eqnarray}
E_{\nu,\rm b}^2\,\Phi_{\nu,\rm b}=\frac{1}{4\pi}\,\Re(0)\,f_z D_{\rm H} \,\frac{3}{8}\,E_{\rm CR}^2\frac{dN_{\rm CR}}{dE}(t_{\nu,\rm b})\, f_{\rm sup}f_{\rm mes}
\end{eqnarray}

where 
%the neutrino luminosity $L_\nu=f_{\rm sup}f_{\rm mes}f_\nu f_\pi L_{\rm CR}$, and 
$D_{\rm H}$ is the Hubble distance corresponding to redshift $z_{\rm H}$, and the factor
\begin{eqnarray}
f_z\equiv\frac{1}{D_{\rm H}}\,\int_0^{z_{\rm H}}\,\frac{1}{1+z}\,\frac{{\rm d}D}{{\rm d}z}\,\frac{\Re(D)}{\Re(0)}\,{\rm d}z \ .
\end{eqnarray}
For a uniform source birthrate $\Re(D)=\Re(0)$, $f_z\sim 0.55$, and for a source emissivity following the star formation rate (SFR) as in Ref.~\cite{2006ApJ...651..142H}, $f_z\sim 2.5$. For magnetars, for which hadronic interactions dominate, the diffuse neutrino flux can then be estimated as  \cite{Murase09} (assuming $t_{\rm sd}\ll t_{\nu, \rm b}$)
\begin{eqnarray}
E_{\nu,\rm b}^2\,\Phi_{\nu,\rm b}&=& 1.5\times 10^{-8}\, {\rm GeV\,cm^{-2}\,s^{-1}\,sr^{-1}} \,\kappa_4^{3/4}\,P_{i, -3}^{-3/4}\,\eta^{3/4}\,B_{15}^{-7/4}\,Z^{-1}\,M_{\rm ej,1}^{-5/8}\nonumber\\ 
&&\times \frac{f_z}{2.25}\,\frac{\Re(0)}{1.2\times 10^3\,\rm Gpc^{-3}\,yr^{-1}}\ .
\end{eqnarray}

We normalize the flux of neutrinos by setting the   associated cosmic-ray flux to the observed level. Assuming an energy loss length $D_{\rm loss}$ on the intergalactic backgrounds at a given energy, the cosmic-ray flux can be estimated roughly as
\begin{eqnarray}
\Phi_{\rm CR}\simeq \frac{1}{4\pi}\,E_{\rm CR}^2\frac{{\rm d}N_{\rm CR}}{{\rm d}E_{\rm CR}}\,\Re(0)\,D_{\rm loss}\ .
\end{eqnarray}
At the cosmic-ray break energy, at time $t_{\rm CR,b}=\max(t_{pp}^*,t_{p\gamma}^*)\sim t_{p\gamma}^*$, the cosmic-ray flux thus reads
\begin{eqnarray}
E_{{\rm CR,b}}^2\,\Phi_{{\rm CR}}&=& 2.0\times 10^{-12}\, {\rm GeV\,cm^{-2}\,s^{-1}\,sr^{-1}} \,AZ^{-1}\eta B_{15}^{-2}\kappa_4\\ \nonumber
&&\eta_{\rm th,0}^{-3/5}P_{i,-3}^{1/5}M_{\rm ej,1}^{-1/2}\frac{\Re(0)}{1.2\times 10^3\,\rm Gpc^{-3}\,yr^{-1}}\,\frac{D_{\rm loss}}{4000\,\rm Mpc}
\end{eqnarray}
In the majority of the parameter-space considered, $E_{{\rm CR},pp}\lesssim 10^{17}\,$eV and we fall in a regime where the energy-loss distance is close to the Hubble distance. Diffusion effects in the intergalactic magnetic fields would alter the distance $D_{\rm loss}$ (see, e.g., \cite{L05,KL08a,KL08b}) significantly for cosmic-ray energies $\gtrsim 10^{17}\,$eV, then we take the corresponding $D_{\rm loss}$ as calculated in \cite{KO11}.

In Figure~\ref{fig:spec} we show the cosmic ray  and neutrino spectra  from a magnetar with $B=10^{15}$ G and $P_i=1$ ms calculated using the above methods. As a consistency check, we also show the spectra calculated via numerical simulations \citep{FKO12, F15}. We find that the two approaches produce comparable results.

The predicted flux is then compared with the observed cosmic ray flux $\Phi_{\rm CR}^{\rm ob}$. In our calculation we take the measurements by KASCADE \citep{Apel:2011mi, Apel:2013ura} and the Auger Observatory \citep{ThePierreAuger:2013eja}. 
Notice that the energy losses by interactions on the cosmic radiation fields during the propagation from the source to the Earth further changes the spectrum. This change is not taken into account here, as the dominant process at these energies are adiabatic losses.

As the main contributors to cosmic rays below $10^{17}\,\rm eV$ are known to be most probably not extragalactic, we thus only consider sources that have the energetics to go above this energy, with $E_0=10^{17}\,\rm eV$. This will be indicated as a green dashed line in the top left corner of our limitting contours  in Sec.~\ref{section:param}.

In addition, we limit the upper bound of the birth rate of the sources to be no more than $20\%$ of supernova birth rate $R_{\rm SN} = 1.2\times10^{5}\,\rm Gpc^{-3}\,yr^{-1}$ \citep{2003ApJ...595..346Z, Murase09}, which can be expressed  
\begin{equation}\label{eqn:rate}
\Re(0)=\min{\left(\frac{\Phi_{\rm CR}^{\rm ob}}{\Phi_{\rm CR}}\,R_0,\,20\% R_{\rm SN}\right)\ .}
\end{equation}

\subsection{Comparison between radiative and hadronic background effects}

\begin{figure}[t]
\centering
\epsfig{file=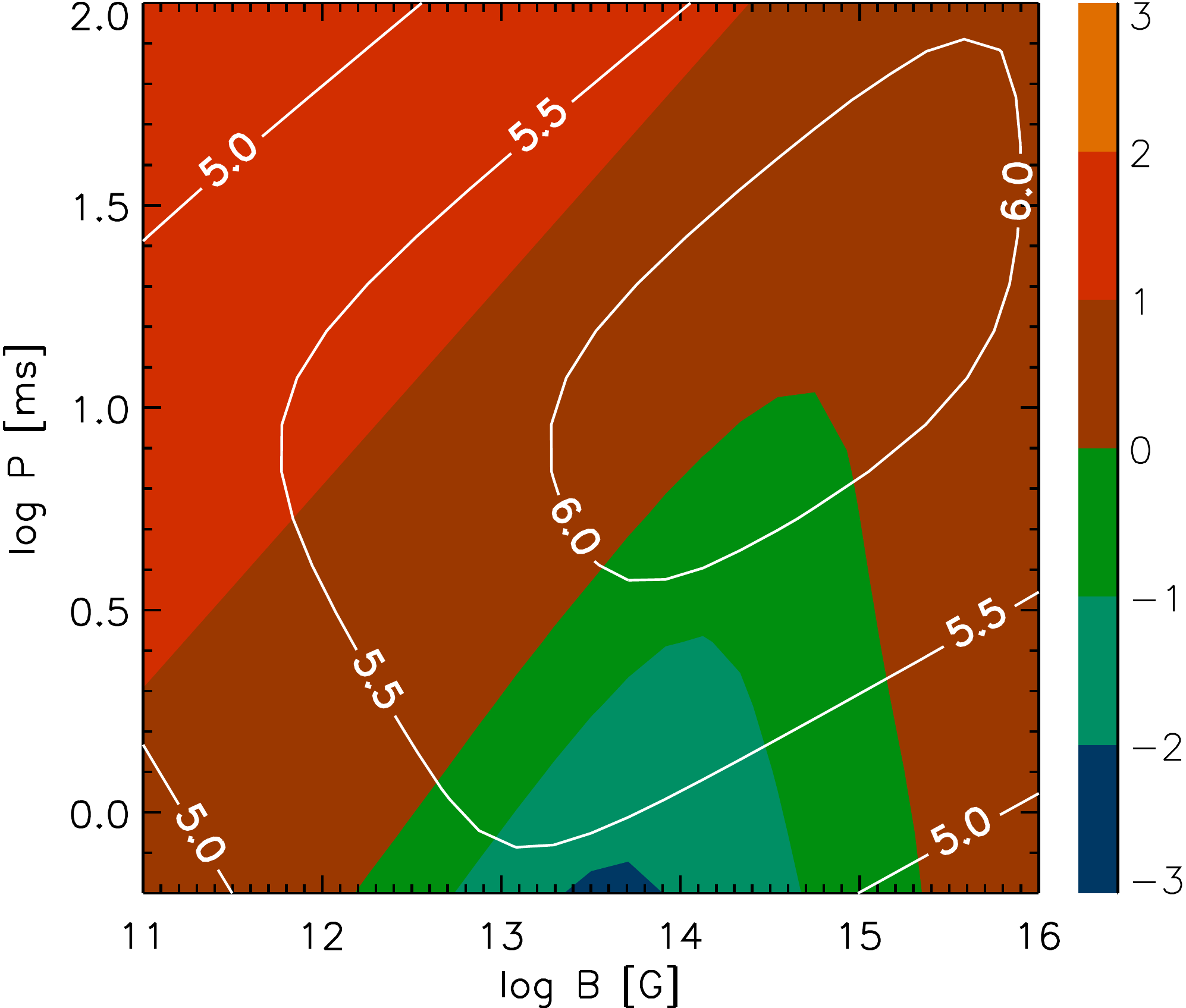,width=0.7\textwidth}
\caption{Interaction timescale ratios $t_{p\gamma}/t_{pp}$ (colored contours) calculated at the break time  $t_{\nu,b}$ defined in Equation~\ref{eqn:t_b} (white contours showing $\log t_{\nu,b}$), for the parameter space $(B,P_{\rm i})$, with $\eta_{\rm th}=1$. The radiation field is dominant over the hadronic backgrounds for the fastest-spinning neutron stars with sub-millisecond periods (green colors). For pulsars with periods more than $\sim 1$ ms, hadronic interactions dominate at $t_{\nu,b}$ (red colors). The lower limit of the y-axis corresponds to the minimum spin period of a neutron star, $P_{i,\rm min}\sim0.6\,\rm ms$ \citep{Haensel99}. }
\label{fig:dominant_proc} 
\end{figure}

As both radiative and hadronic backgrounds evolve with time, the dominant process in a  pulsar also changes over time. The time $t_{\nu,b}$ when neutrino spectrum breaks due to the ending of $\pi p$ or $\pi\gamma$ suppression (defined in Equation~\ref{eqn:t_b}) serves as a good reference time for the system. 
Figure~\ref{fig:dominant_proc} presents the ratios $t_{pp}/t_{p\gamma}$ (colored contours) calculated at $t_{\nu,b}$ (white contours showing $\log t_{\nu,b}$), for the parameter space $(B,P_{\rm i})$, with $\eta_{\rm th}=1$.  The radiation field  is dominant  over the hadronic backgrounds for the fastest-spinning neutron stars with spin period less than $\sim$ 1 ms.  This is because the higher rotational speed the star has, the more energy can be converted to the thermal photons. 
 To have hadronic interactions always  dominate over photopion interactions over the entire $(P,B)$ parameter-space studied here, $\eta_{\rm th}$ needs to be as low as $\sim 10^{-3}$.  Our  calculation here does consider super-luminous supernova. For super-luminous cases, the radiation energy can reach $ 10^{51}$ erg, and the $p\gamma$ process  would be more important.

Note however that this calculation assumes that the radiation background   in the supernova ejecta is isotropic. In the case of a jet-like structure, the radiation field would be beamed and the corresponding photon energy experienced by the proton would scale with $1/\Gamma$, $\Gamma$ being the Lorentz factor of the jet. For lower dissipation efficiency into radiation and beamed radiation, the contribution of the radiative background should be lower unless jet emission is intense. It is thus likely that the hadronic interactions are dominant over a large fraction of the parameter space that enables the acceleration of particles to UHE.  This case is discussed in Section~\ref{sec:jet}.

One caveat of this comparison  assumes that accelerated cosmic rays undergo interactions with radiative and baryonic backgrounds at the same time. However in a realistic picture, cosmic rays would most probably interact with the thermal photons that fill the entire pulsar wind bubble first, and then with baryons when they manage to go through the bubble. In that case, the region filled with warm color in Fig.~\ref{fig:dominant_proc} should still be significantly impacted by the photopion process.

\section{Parameter scan and the viable neutron stars}\label{section:param}

\begin{figure}[h]
\centering
\epsfig{file=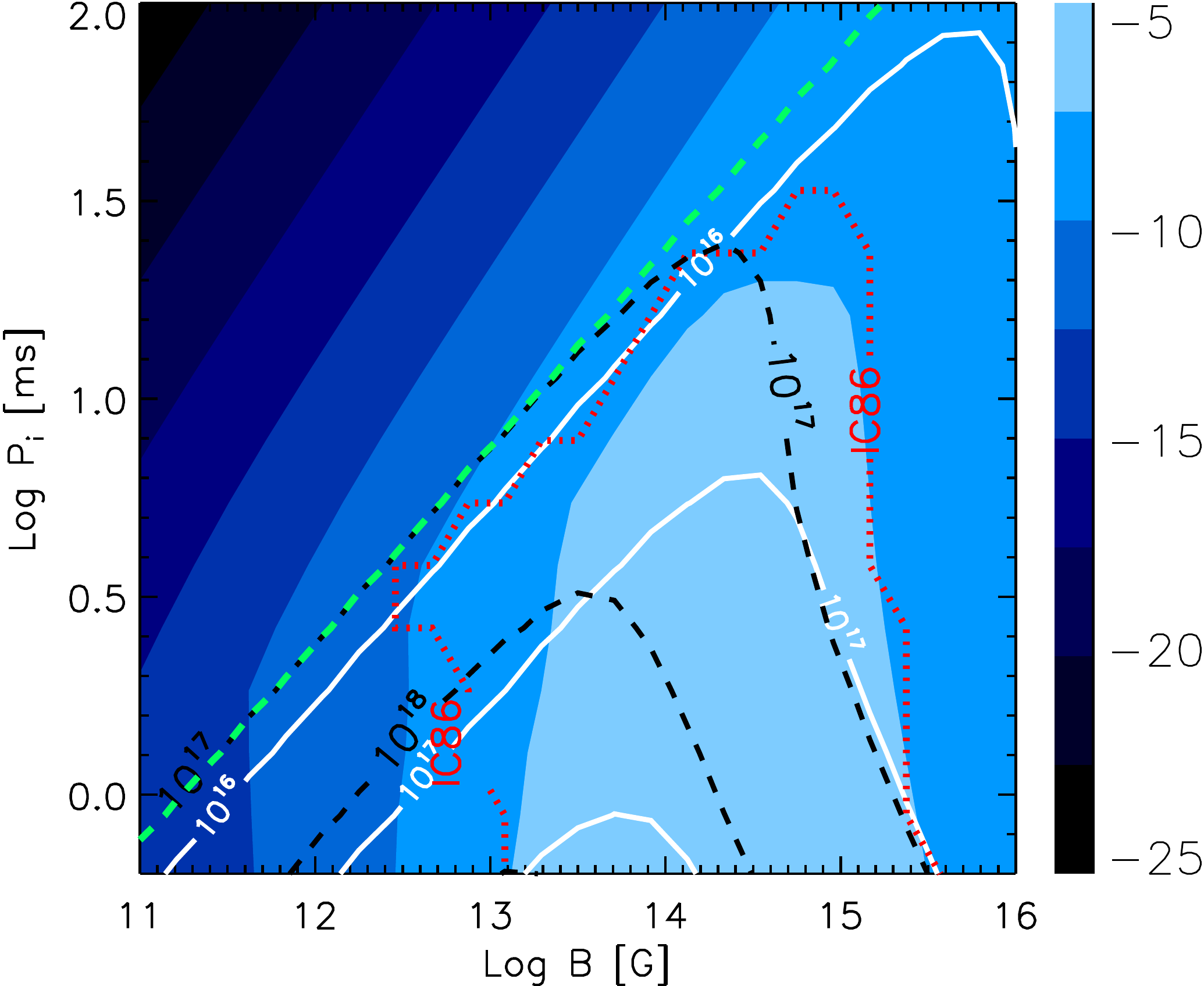,width=0.7\textwidth}
\caption{The neutrino flux $\log_{10}({E_{\nu,\rm b}^2\Phi_{\nu,\rm b}})$ in $\rm GeV\,cm^{-2}\,s^{-1}\,sr^{-1}$  emitted by populations of neutron stars with the same characteristic $(B,P)$, assuming acceleration efficiency $\eta=1$,  ejecta mass $M_{\rm ej}=10\,M_{\odot}$, jet fraction $f_{\rm jet}=0$ and source emissivity following a SFR evolution. The sources birthrates are normalized via cosmic ray measurements. Only hadronic backgrounds is considered for the interactions. Overlaid are the IceCube 5-year sensitivity limit \citep{Abbasi:2011zx, 2013ApJ...779..132A} (red dotted), cosmic ray peak energies $E_{\rm CR,b}$ (black dashed), and neutrino break energies $E_{\nu,{\rm b}}$ (white).  We only consider parameter region below the green dashed line, which encloses sources that can produce cosmic rays above $10^{17}\,\rm eV$.  The parameter space   below the red dotted line is excluded. 
}
\label{fig:neutflux_nojet_hadron} 
\end{figure}

\begin{figure}[h]
\centering
\epsfig{file=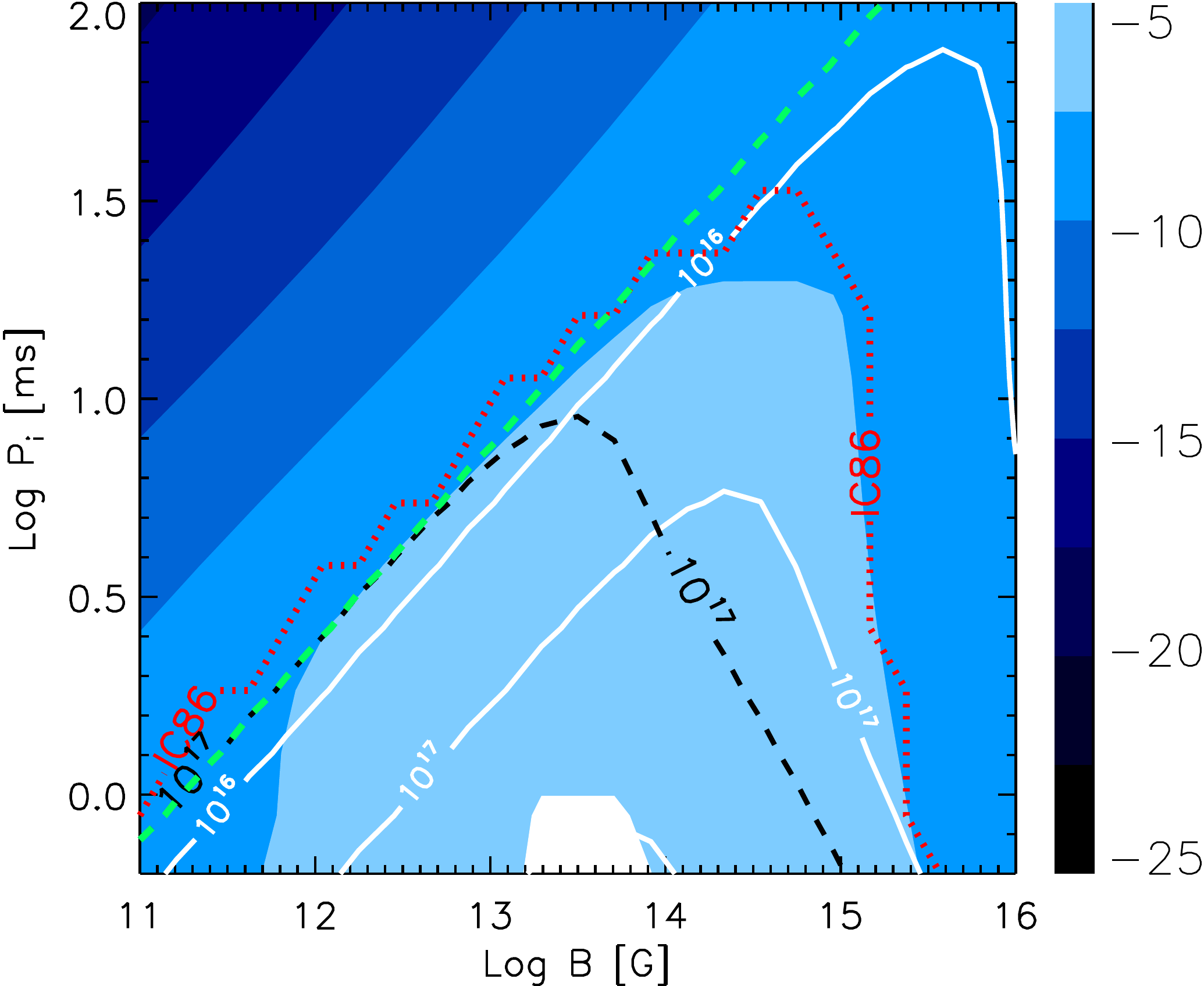,width=0.7\textwidth}
\caption{Same as Fig.~\ref{fig:neutflux_nojet_hadron}, but considering both hadronic and radiative backgrounds. For the radiative background, we assume $\eta_{\rm th}=1$, and the thermal  component dominants over the non-thermal component (see Sec.~\ref{sec:radiation} for details). }
\label{fig:neutflux_nojet} 
\end{figure}

\begin{figure}[h]
\centering
\epsfig{file=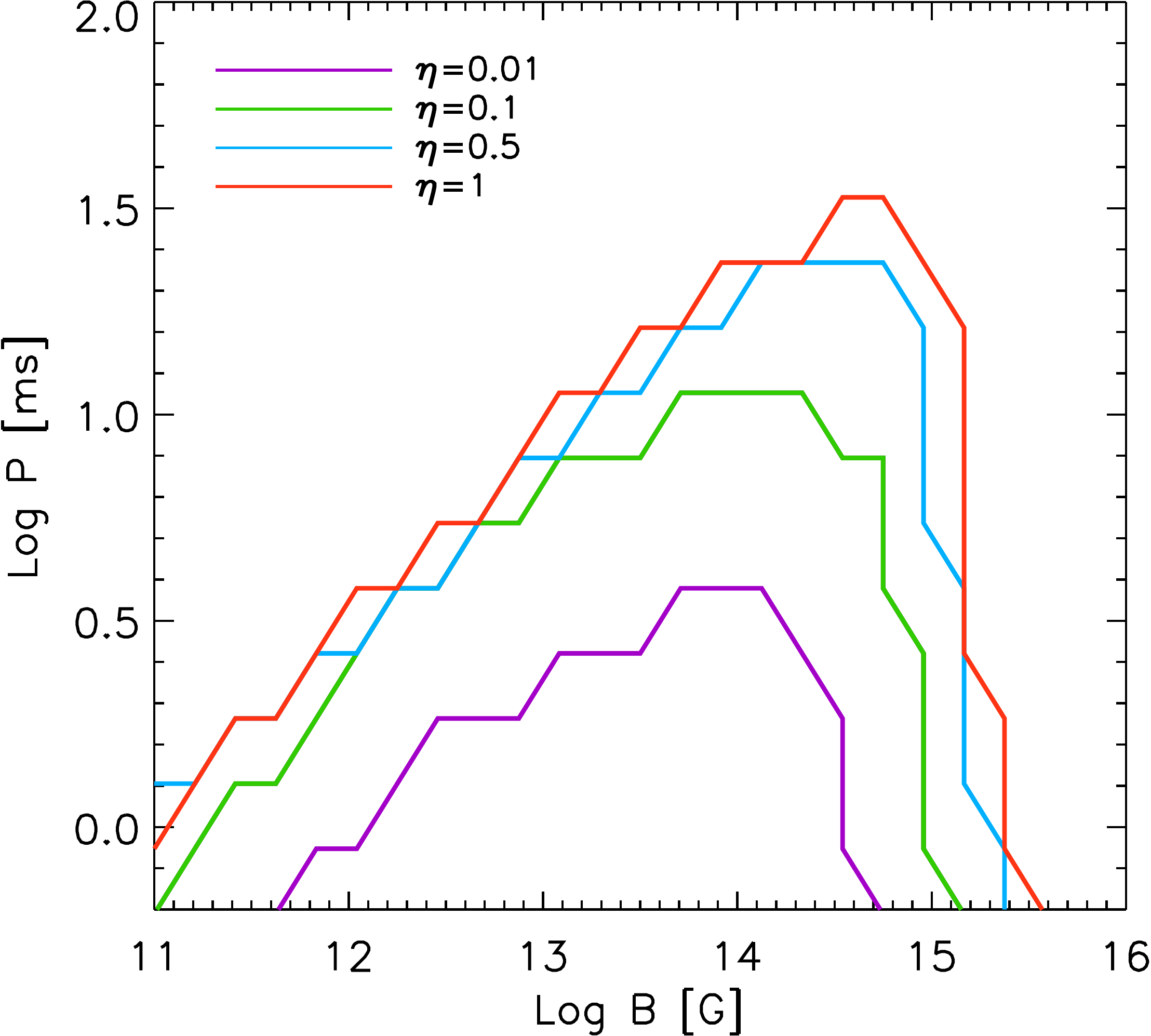,width=0.7\textwidth}
\caption{Limiting contours for different acceleration efficiency $\eta =0.01, 0.1, 0.5, 1$. As in the previous plots, the parameter space   below  the contours is excluded by IceCube. All cases assume ejecta mass $M_{\rm ej}=10\,M_{\odot}$, no jet and source emissivity following SFR evolution. The sources birthrates are normalized via cosmic ray measurements. Both hadronic backgrounds and radiative backgrounds assuming $\eta_{\rm th}=1$ are considered for the interactions.}
\label{fig:multieta} 
\end{figure}

Figures~\ref{fig:neutflux_nojet_hadron} and \ref{fig:neutflux_nojet} show the expected neutrino flux $\log_{10}({E_{\nu,\rm b}^2\Phi_{\nu,\rm b}})$ emitted by populations of neutron stars over the parameter space of $(B, P)$,  considering only hadronic interactions (Fig.~\ref{fig:neutflux_nojet_hadron}) and both hadronic and photo-pion interactions (Fig.~\ref{fig:neutflux_nojet}). In this parameter scan we assume that  neutron stars have an acceleration efficiency $\eta=1$, ejecta mass $M_{\rm ej} =10\,M_\odot$, and a source emissivity following a uniform distribution. In these plots we have calibrated the source birthrate using cosmic ray observations. Specifically, for each point in the parameter space, the birth rate of the neutron star population with characteristics $(P,B)$ is calculated so as to fit the measured cosmic ray flux at $E_{\rm CR,b}$, following Eq.~(\ref{eqn:rate}). 

We first consider hadronic interactions in Fig.~\ref{fig:neutflux_nojet_hadron}. 
The white contours present the neutrino break energy $E_{\nu,\rm b}$  where the neutrino spectrum peaks, and the green contours present the peak energies $E_{\rm CR,b}$ of high-energy cosmic rays from the neutron stars.  $E_{\nu,\rm b}$ reaches  $10^{18}\,\rm eV$ in the parameter region of $10^{12}\,\rm G < B < 10^{15}\,\rm G$ and $P_{i} < 0.6\,\rm ms$.  For larger $P$ and smaller $B$, $E_{\nu,\rm b}$ decreases because the star is less energetic. On the other hand, when $B>10^{14}\,\rm G$, magnetars have a fast spin-down time while the system remains too opaque for pions to decay, which results a cutoff on the neutrino peak energy.  The dotted red line delimits the region where neutron stars would emit a diffusive neutrino background that exceeds the 5-year sensitivity of the IceCube Observatory.

Figure~\ref{fig:neutflux_nojet} depicts the same parameter region, but additionally takes into account the radiation background in the pulsar winds. A thermalization factor of $\eta_{\rm th} = 1$ is assumed for this calculation. The radiation background does not change the neutrino break energy  significantly. However, as the radiation field decreases much slower than the hadron background, it causes the cosmic ray spectrum to break much later in time compared to the hadronic case. As a result, the lower cosmic ray flux at  $E_{\rm CR,b}$   implies a higher normalization for the neutrino flux, and a larger parameter region is constrained by observations.

In Fig.~\ref{fig:multieta} we present the limit contours for a range of acceleration efficiency with $\eta = [0.01, 0.1, 0.5, 1]$. For all four cases, we assume ejecta mass $M_{\rm ej}=10\,M_\odot$, no jet configuration, and source emissivity following SFR. Both hadronic and radiation backgrounds with $\eta_{\rm th}=1$ are considered.  We find no strong variation  in cases with $\eta\geq0.5$. For  $\eta<0.5$, the confinement area decreases for smaller $\eta$. Note that in the case $\eta=0.01$, the maximum cosmic ray energy accelerated by the pulsar wind can only reach $10^{18}\,$eV.

\section{In presence of jet-like structures}\label{sec:jet}

\begin{figure}[t]
\centering
\epsfig{file=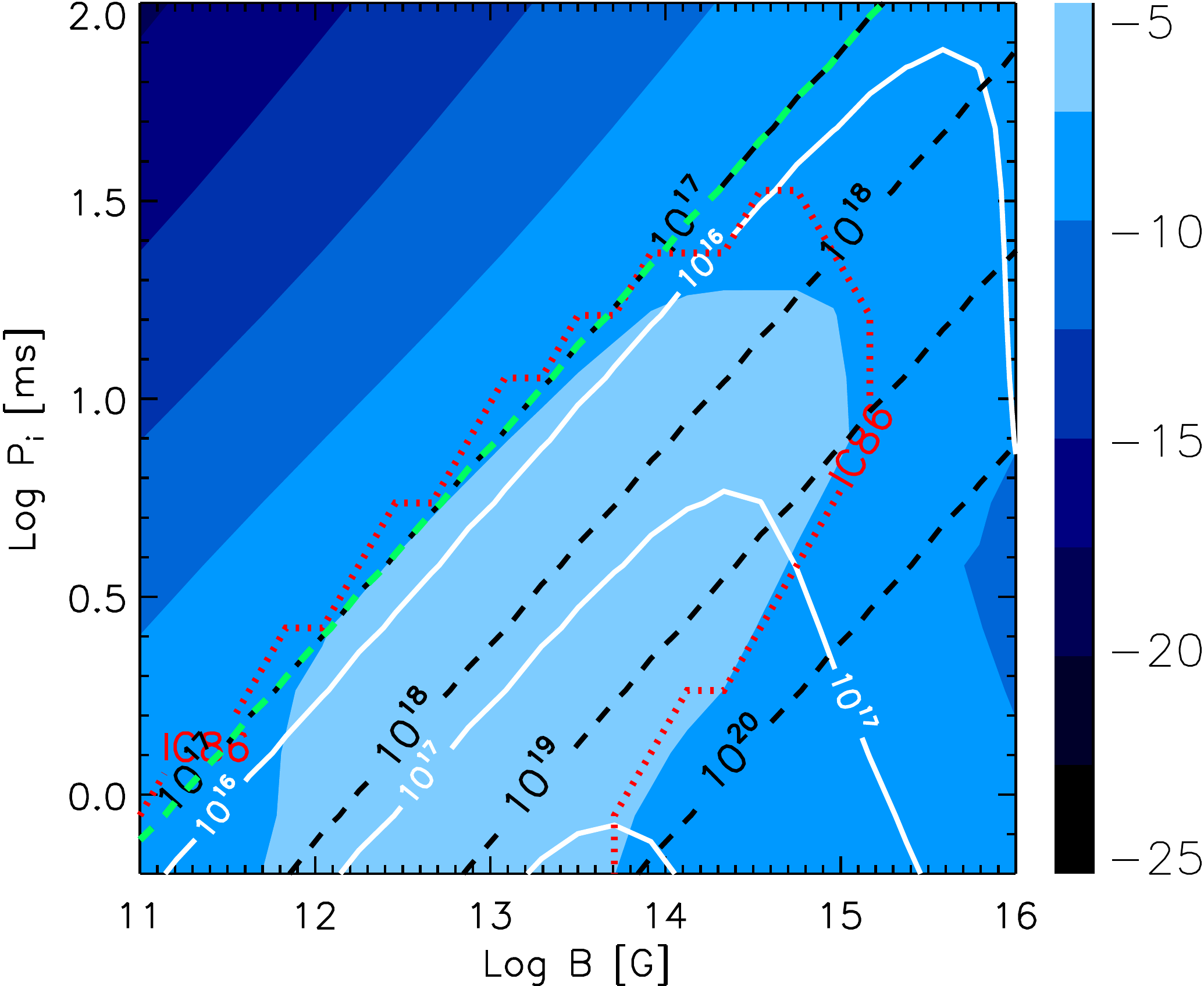,width=0.7\textwidth}
\caption{The neutrino flux $\log_{10}({E_{\nu,\rm b}^2\Phi_{\nu,\rm b}})$ in $\rm GeV\,cm^{-2}\,s^{-1}\,sr^{-1}$  emitted by populations of neutron stars with the same characteristic $(B,P)$, assuming acceleration efficiency $\eta=1$,  ejecta mass $M_{\rm ej}=10\,M_{\odot}$, jet fraction $f_{\rm jet}=0.1$ and source emissivity following SFR. The sources birthrates are normalized via cosmic ray measurements. Both hadronic backgrounds and radiative backgrounds assuming $\eta_{\rm th}=1$ are considered for the interactions. Overplotted are the IceCube sensitivity limit (red dotted), cosmic ray peak energie $E_{\rm cr, peak}$ (black dashed), and neutrino break energies $E_{\nu,{\rm b}}$ (white).  The parameter space within the red dotted line is excluded.}
\label{fig:neutflux_jet} 
\end{figure}

\begin{figure}[t]
\centering
\epsfig{file=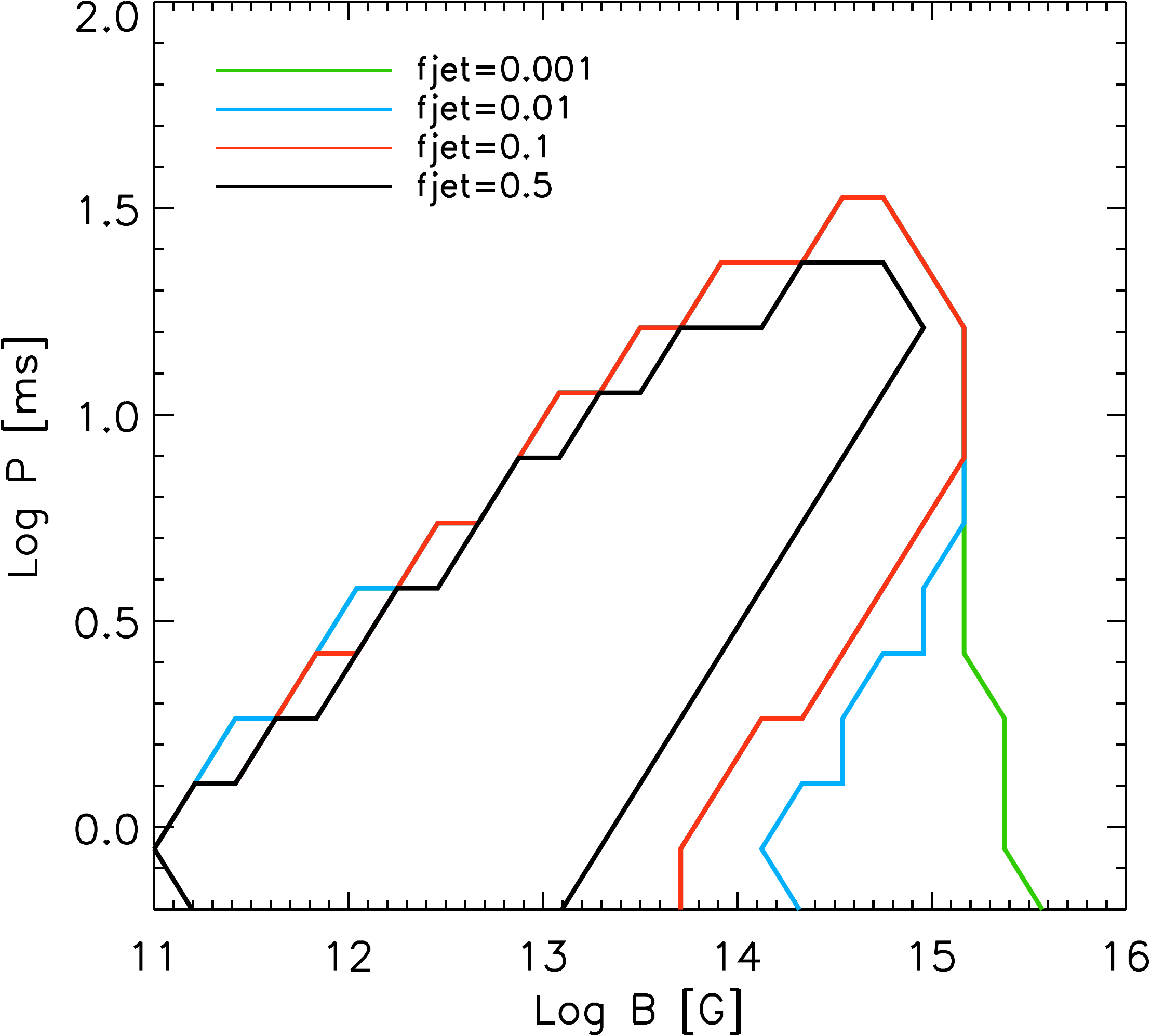,width=0.7\textwidth}
\caption{Limiting contours for different jet fraction $f_{\rm jet} = 0.001, 0.01, 0.1, 0.5$. Like previous plots, the parameter space within the contours is excluded by IceCube. All cases assume ejecta mass $M_{\rm ej}=10\,M_{\odot}$, $\eta=1$ and source emissivity following SFR. The sources birthrates are normalized via cosmic ray measurements. Both hadronic backgrounds and radiative backgrounds assuming $\eta_{\rm th}=1$ are considered for the interactions.   }
\label{fig:multijet} 
\end{figure}

The confining pressure of the toroidal magnetic field could collimate the proto-magnetar wind along its polar axis, and drive a jet that has the properties of long gamma-ray bursts jets \citep{Usov92, 2004ApJ...611..380T, Komissarov07,Bucciantini07,Bucciantini08,Bucciantini09}. Cosmic rays accelerated inside the proto-magnetar jet could then escape through the pierced supernova envelope. The escape of nuclei through jets, taking into account the radiative and baryonic background fields, has been studied numerically and semi-analytically in the context of GRBs by Ref.~\cite{Murase08} and for proto-magnetar jets by Ref.~\cite{Metzger11}. 

The collimation power becomes significant for values of the ratio of the Poynting flux to the total energy at the termination shock of the wind, $\dot E_{\rm mag}/\dot E_{\rm tot} \gtrsim 0.2$, at times $t\sim 10-100$~s~\cite{Bucciantini07}. The conversion of magnetic energy into kinetic energy in relativistic outflows at large radii are uncertain and might not allow the formation of a jet. Studies of the Crab Pulsar wind nebula show indeed that $\dot E_{\rm mag}/\dot E_{\rm tot}\sim 10^{-2}$ at large radii \citep{Kennel84,Begelman92}, but magnetars could have different fates. 

Reference~\cite{Arons03} also proposed that the supernova ejecta could be disrupted by the magnetar wind. Such phenomena have never been observed, neither in magnetar envelopes, nor in rotation-powered pulsar envelopes. 

We parametrize the uncertainties of these escape scenarios by introducing the quantity $f_{\rm jet}$, that gives the fraction of accelerated particles that can escape without crossing a dense environment. Note that in the jet scenario we still assume that all particles get accelerated in the pulsar magnetosphere, so  particles injected off the jet-axis will still undergo interactions with the rest of the ejecta that has not been pierced. 

The flux of cosmic rays escaped from jets is  compared to the observed cosmic ray flux, putting an extra constraint on the magnetar  birth rate, in addition to that from the  cosmic rays leave  from the non-jet region after interactions:
\begin{equation}
\Re(0)=\min{\left(\frac{\Phi_{\rm CR}^{\rm ob}}{\Phi_{\rm CR,sp}}R_0,\, \frac{\Phi_{\rm CR}^{\rm ob}}{\Phi_{\rm CR,jet}}\,R_0, 20\%\,R_{\rm SN}\right)}
\end{equation}
where ${\Phi_{\rm CR,jet}} = f_{\rm jet}\,E_0^2\,dN/dE\Re(0)D_{\rm loss}/4\pi$ is the cosmic ray flux from the jet region peaking at $E_0$, while $\Phi_{\rm CR, sp} = (1-f_{\rm jet})\,\Phi_{\rm CR}$ is the flux from the non-jet region peaking at $E_{\rm CR,b}$. 

Figure~\ref{fig:neutflux_jet} presents the neutrino flux over the parameter space assuming a jet fraction $f_{\rm jet}=0.1$. The parameter region with $B>10^{14}\,\rm G$ and $P<10\,\rm ms$ is significantly impacted. The reason is that the jet allows the escape of UHECRs accelerated by the fast-spinning magnetars in this region, which poses a strong limit on the star burst rate due to their low flux. The constraints on magnetars with larger spin periods still hold.

Figure~\ref{fig:multijet} shows the limiting contours for different jet fractions with $f_{\rm jet} = [0.001, 0.01, 0.1, 0.5]$. $\eta=1$ and source emissivity following SFR have been assumed for all cases. As expected, when $f_{\rm jet}\ll 1$, the results get back to cases without any jet configuration. The parameter space is less confined with a large $f_{\rm jet}$, as most cosmic rays escape without producing neutrinos. Interestingly, the most standard magnetars  remain excluded even with a large jet fraction.

\section{Discussion, conclusions} \label{section:conclusion}
In this work we have constrained magnetars as sources of cosmic rays above $10^{17}\,\rm eV$, by comparing their expected neutrino production to the observational limits measured by the IceCube Observatory. We have considered particle interactions with both radiative field in the pulsar wind nebula and the hadronic backgrounds of the supernova ejecta. High-energy neutrinos provide a powerful tool to probe very high-energy cosmic-ray acceleration hidden in supernova ejecta \cite{Murase09}. Assuming a proton cosmic ray composition, we find that the assumption of magnetars being the dominant high-energy cosmic ray sources is mostly ruled out by the IceCube upper limits on the diffusive neutrino background,  unless the ejecta mass is much smaller than in a typical core-collapse supernova, or a large fraction of cosmic rays can escape without significant interactions from the jet-like structures piercing the ejecta. 

Throughout the work we have assumed that cosmic rays are mostly composed of protons.  If cosmic rays are instead nuclei with mass number A, their meson production efficiency from interactions with hadronic backgrounds would decrease by $f_{{\rm mes}, Np}\propto A^{-1/3}$ \citep{FKMO14_letter}, while that from the photodisintegration drops to $f_{\rm mes, p\gamma}\propto A^{-1.21}$ \citep{Murase:2010gj}. On the other hand, as Eq.~(\ref{eqn:Et}) shows, nuclei would be accelerated at much later time than protons at the same energy. As a result, the environment would be less   dense and particles could escape with higher energies compared to the proton scenario. In the end  we would expect less neutrino production and a less constrained parameter space if cosmic rays have a heavy composition.  However, we note that a large number of higher-order products, including secondary mesons and neutrinos, would also  result from the $Np$ and $N\gamma$ interactions and thus help to constrain the parameter space. Indeed, as Ref.~\cite{FKMO14_letter} shows, there is no significant difference between the neutrino flux produced by protons and iron nuclei primaries, if considering only hadronic interactions. 

In Section~\ref{sec:jet}, we showed that the presence of jets in magnetars enabling the escape of UHECRs can help remove the tension between their EeV neutrino productions and the observation. In particular, we  demonstrate that if $10\%$ cosmic rays can leak from the jet structure, fast-spinning magnetars   shift outside of the exclusion region due to the IceCube limits, as the low UHECR flux ensures a rare magnetar birth rate.  

With equation~\ref{eqn:rate} we have assumed that the birth rate of the sources is no more than $20\%$ of normal supernova birth rate. This upper limit could be over-estimated for fast-spinning magnetars. \cite{2010ApJ...717..245K} showed that fast-spinning strongly magnetized pulsars could lead to  superluminous supernovae, which are found to be as rare as $0.01\%-0.1\%$ normal core collapse events \citep{2011Natur.474..487Q}. A much lower birth rate could help remove the constraints on fast-spinning magnetars as sources of high-energy neutrinos, but the conclusion that magnetars cannot be the dominant sources of  cosmic rays above $10^{17}$ eV still holds.

Reference \cite{Faucher06} suggested that neutron stars at birth have spin periods  that follow a normal distribution with mean $300\,\rm ms$ and standard derivation $150\,\rm ms$, and the log of their dipole magnetic fields that follow a normal distribution with mean $12.65$ and standard derivation $0.55$. Here we have ignored the effect of such a distribution as we aim to separate the contribution from different parts of the parameter space.  In particular, we focus on magnetars in this work, and it is not obvious that these objects present such a distribution. Cosmic ray and neutrino productions from a cumulation of  neutron star population following a $(P, B)$ distribution can be found in  \cite{FKO13, FKMO14_letter}.  We stress that the conclusions of these works (that focus on mildly magnetized pulsars) are not in contradiction with the present paper, because of the population distribution and the injection of a non-proton composition.

\acknowledgments
KF acknowledges financial support from the JSI fellowship at the University of Maryland. 
KK acknowledges financial support from the PER-SU fellowship at Sorbonne Universit\'es and from the Labex ILP (reference ANR-10-LABX-63, ANR-11-IDEX-0004-02).
KM acknowledges Institute for Advanced Study for continuous support.
AO acknowledge financial support from the NSF grant NSF PHY-1412261 and the NASA grant 11-APRA-0066 at the University of Chicago, and the grant NSF PHY-1125897 at the Kavli Institute for Cosmological Physics.

\bibliography{FKMO15}

\providecommand{\href}[2]{#2}\begingroup\raggedright\begin{thebibliography}{10}

\bibitem{Cesarsky80}
C.~J. Cesarsky, {\it Cosmic-ray confinement in the galaxy},  {\em Ann. Rev.
  Astron. Astrophys.} {\bf 18} (1980), no.~1 289--319,
  [\href{http://xxx.lanl.gov/abs/http://www.annualreviews.org/doi/pdf/10.1146/annurev.aa.18.090180.001445}{{\tt
  http://www.annualreviews.org/doi/pdf/10.1146/annurev.aa.18.090180.001445}}].

\bibitem{Hillas84}
A.~M. {Hillas}, {\it {The Origin of Ultra-High-Energy Cosmic Rays}},  {\em
  ARAA} {\bf 22} (1984) 425--444.

\bibitem{Strong07}
A.~W. Strong, I.~V. Moskalenko, and V.~S. Ptuskin, {\it {Cosmic-ray propagation
  and interactions in the Galaxy}},  {\em Ann. Rev. Nucl. Part. Sci.} {\bf 57}
  (2007) 285--327, [\href{http://xxx.lanl.gov/abs/astro-ph/0701517}{{\tt
  astro-ph/0701517}}].

\bibitem{Tavani10}
M.~{Tavani}, A.~{Giuliani}, A.~W. {Chen}, et~al., {\it {Direct Evidence for
  Hadronic Cosmic-Ray Acceleration in the Supernova Remnant IC 443}},  {\em ApJ
  Lett.} {\bf 710} (Feb., 2010) L151--L155,
  [\href{http://xxx.lanl.gov/abs/1001.5150}{{\tt arXiv:1001.5150}}].

\bibitem{Giuliani11}
A.~{Giuliani}, M.~{Cardillo}, M.~{Tavani}, Y.~{Fukui}, et~al., {\it {Neutral
  Pion Emission from Accelerated Protons in the Supernova Remnant W44}},  {\em
  ApJ Lett.} {\bf 742} (Dec., 2011) L30--L35,
  [\href{http://xxx.lanl.gov/abs/1111.4868}{{\tt arXiv:1111.4868}}].

\bibitem{Ackermann13}
M.~{Ackermann}, M.~{Ajello}, A.~{Allafort}, et~al., {\it {Detection of the
  Characteristic Pion-Decay Signature in Supernova Remnants}},  {\em Science}
  {\bf 339} (Feb., 2013) 807--811,
  [\href{http://xxx.lanl.gov/abs/1302.3307}{{\tt arXiv:1302.3307}}].

\bibitem{Giacinti11}
G.~Giacinti, M.~Kachelriess, D.~Semikoz, and G.~Sigl, {\it {Ultrahigh Energy
  Nuclei in the Turbulent Galactic Magnetic Field}},  {\em Astropart.Phys.}
  {\bf 35} (2011) 192--200, [\href{http://xxx.lanl.gov/abs/1104.1141}{{\tt
  arXiv:1104.1141}}].

\bibitem{Abreu12}
{Pierre Auger Collaboration}, P.~{Abreu}, M.~{Aglietta}, M.~{Ahlers}, E.~J.
  {Ahn}, I.~F.~M. {Albuquerque}, D.~{Allard}, I.~{Allekotte}, J.~{Allen},
  P.~{Allison}, and et~al., {\it {Search for Point-like Sources of Ultra-high
  Energy Neutrinos at the Pierre Auger Observatory and Improved Limit on the
  Diffuse Flux of Tau Neutrinos}},  {\em ApJ Lett.} {\bf 755} (Aug., 2012) L4,
  [\href{http://xxx.lanl.gov/abs/1210.3143}{{\tt arXiv:1210.3143}}].

\bibitem{Abu-Zayyad13}
T.~{Telescope Array}, {Pierre Auger Collaborations}, {:}, T.~{Abu-Zayyad},
  M.~{Allen}, R.~{Anderson}, R.~{Azuma}, E.~{Barcikowski}, J.~W. {Belz}, D.~R.
  {Bergman}, and et~al., {\it {Pierre Auger Observatory and Telescope Array:
  Joint Contributions to the 33rd International Cosmic Ray Conference (ICRC
  2013)}},  {\em ArXiv e-prints} (Oct., 2013)
  [\href{http://xxx.lanl.gov/abs/1310.0647}{{\tt arXiv:1310.0647}}].

\bibitem{KO11}
K.~Kotera and A.~V. Olinto, {\it {The Astrophysics of Ultrahigh Energy Cosmic
  Rays}},  {\em Ann.Rev.Astron.Astrophys.} {\bf 49} (2011) 119--153,
  [\href{http://xxx.lanl.gov/abs/1101.4256}{{\tt arXiv:1101.4256}}].

\bibitem{Venkatesan97}
A.~{Venkatesan}, M.~C. {Miller}, and A.~V. {Olinto}, {\it {Constraints on the
  Production of Ultra--High-Energy Cosmic Rays by Isolated Neutron Stars}},
  {\em ApJ} {\bf 484} (July, 1997) 323--+,
  [\href{http://xxx.lanl.gov/abs/astro-ph/}{{\tt astro-ph/}}].

\bibitem{Blasi00}
P.~{Blasi}, R.~I. {Epstein}, and A.~V. {Olinto}, {\it {Ultra-High-Energy Cosmic
  Rays from Young Neutron Star Winds}},  {\em ApJ Letters} {\bf 533} (Apr.,
  2000) L123--L126.

\bibitem{Arons03}
J.~{Arons}, {\it {Magnetars in the Metagalaxy: An Origin for Ultra-High-Energy
  Cosmic Rays in the Nearby Universe}},  {\em ApJ} {\bf 589} (June, 2003)
  871--892.

\bibitem{Murase09}
K.~{Murase}, P.~{M{\'e}sz{\'a}ros}, and B.~{Zhang}, {\it {Probing the birth of
  fast rotating magnetars through high-energy neutrinos}},  {\em Phys.~Rev.~D}
  {\bf 79} (May, 2009) 103001--+,
  [\href{http://xxx.lanl.gov/abs/0904.2509}{{\tt arXiv:0904.2509}}].

\bibitem{K11}
K.~{Kotera}, {\it {Ultrahigh energy cosmic ray acceleration in newly born
  magnetars and their associated gravitational wave signatures}},  {\em Phys.
  Rev. D} {\bf 84} (July, 2011) 023002,
  [\href{http://xxx.lanl.gov/abs/1106.3060}{{\tt arXiv:1106.3060}}].

\bibitem{FKO12}
K.~{Fang}, K.~{Kotera}, and A.~V. {Olinto}, {\it {Newly Born Pulsars as Sources
  of Ultrahigh Energy Cosmic Rays}},  {\em The Astrophysical Journal} {\bf 750}
  (May, 2012) 118, [\href{http://xxx.lanl.gov/abs/1201.5197}{{\tt
  arXiv:1201.5197}}].

\bibitem{FKO13}
K.~{Fang}, K.~{Kotera}, and A.~V. {Olinto}, {\it {Ultrahigh energy cosmic ray
  nuclei from extragalactic pulsars and the effect of their Galactic
  counterparts}},  {\em JCAP} {\bf 3} (Mar., 2013) 10,
  [\href{http://xxx.lanl.gov/abs/1302.4482}{{\tt arXiv:1302.4482}}].

\bibitem{Lorimer08}
D.~R. Lorimer, {\it Binary and millisecond pulsars},  {\em Living Reviews in
  Relativity} {\bf 11} (2008), no.~8.

\bibitem{Murase:2008sa}
K.~Murase and H.~Takami, {\it {Implications of Ultra-High-Energy Cosmic Rays
  for Transient Sources in the Auger Era}},  {\em Astrophys. J.} {\bf 690}
  (2009) L14--L17, [\href{http://xxx.lanl.gov/abs/0810.1813}{{\tt
  arXiv:0810.1813}}].

\bibitem{Katz09}
B.~{Katz}, R.~{Budnik}, and E.~{Waxman}, {\it {The energy production rate and
  the generation spectrum of UHECRs}},  {\em Journal of Cosmology and
  Astro-Particle Physics} {\bf 3} (Mar., 2009) 20--+,
  [\href{http://xxx.lanl.gov/abs/0811.3759}{{\tt arXiv:0811.3759}}].

\bibitem{Woods06}
P.~M. {Woods} and C.~{Thompson}, {\em {Soft gamma repeaters and anomalous X-ray
  pulsars: magnetar candidates}}, pp.~547--586.
\newblock In: Compact stellar X-ray sources. Edited by Walter Lewin \& Michiel
  van der Klis. Cambridge Astrophysics Series, No. 39. Cambridge, UK: Cambridge
  University Press, ISBN 978-0-521-82659-4, Apr., 2006.

\bibitem{Harding06}
A.~K. {Harding} and D.~{Lai}, {\it {Physics of strongly magnetized neutron
  stars}},  {\em Reports on Progress in Physics} {\bf 69} (Sept., 2006)
  2631--2708, [\href{http://xxx.lanl.gov/abs/astro-ph/}{{\tt astro-ph/}}].

\bibitem{Mereghetti08}
S.~{Mereghetti}, {\it {The strongest cosmic magnets: soft gamma-ray repeaters
  and anomalous X-ray pulsars}},  {\em The Astronomy and Astrophysics Review}
  {\bf 15} (July, 2008) 225--287,
  [\href{http://xxx.lanl.gov/abs/0804.0250}{{\tt arXiv:0804.0250}}].

\bibitem{Duncan92}
R.~C. {Duncan} and C.~{Thompson}, {\it {Formation of very strongly magnetized
  neutron stars - Implications for gamma-ray bursts}},  {\em ApJ Letters} {\bf
  392} (June, 1992) L9--L13.

\bibitem{Kouveliotou98}
C.~{Kouveliotou}, S.~{Dieters}, T.~{Strohmayer}, J.~{van Paradijs}, G.~J.
  {Fishman}, C.~A. {Meegan}, K.~{Hurley}, J.~{Kommers}, I.~{Smith}, D.~{Frail},
  and T.~{Murakami}, {\it {An X-ray pulsar with a superstrong magnetic field in
  the soft {$\gamma$}-ray repeater SGR1806 - 20}},  {\em Nature} {\bf 393}
  (May, 1998) 235--237.

\bibitem{Kouveliotou99}
C.~{Kouveliotou}, T.~{Strohmayer}, K.~{Hurley}, J.~{van Paradijs}, M.~H.
  {Finger}, S.~{Dieters}, P.~{Woods}, C.~{Thompson}, and R.~C. {Duncan}, {\it
  {Discovery of a Magnetar Associated with the Soft Gamma Repeater SGR
  1900+14}},  {\em ApJ Letters} {\bf 510} (Jan., 1999) L115--L118,
  [\href{http://xxx.lanl.gov/abs/astro-ph/}{{\tt astro-ph/}}].

\bibitem{Baring01}
M.~G. {Baring} and A.~K. {Harding}, {\it {Photon Splitting and Pair Creation in
  Highly Magnetized Pulsars}},  {\em ApJ} {\bf 547} (Feb., 2001) 929--948,
  [\href{http://xxx.lanl.gov/abs/astro-ph/}{{\tt astro-ph/}}].

\bibitem{AugerIcrc11}
P.~{Abreu} et~al., {\it The pierre auger observatory ii: Studies of cosmic ray
  composition and hadronic interaction models},  {\em 32nd International Cosmic
  Ray Conference, Beijing, China} (2011) arXiv:1107.4804.

\bibitem{TAicrc11}
Y.~{Tameda} et~al. {\em 32nd International Cosmic Ray Conference, Beijing,
  China} {\bf 2} (2011), no.~246.

\bibitem{FKMO14_letter}
K.~{Fang}, K.~{Kotera}, K.~{Murase}, and A.~V. {Olinto}, {\it {A decisive test
  for the young pulsar origin of ultrahigh energy cosmic rays with IceCube}},
  {\em ArXiv e-prints} (Nov., 2013)
  [\href{http://xxx.lanl.gov/abs/1311.2044}{{\tt arXiv:1311.2044}}].

\bibitem{Ruderman75}
M.~A. {Ruderman} and P.~G. {Sutherland}, {\it {Theory of pulsars - Polar caps,
  sparks, and coherent microwave radiation}},  {\em ApJ} {\bf 196} (Feb., 1975)
  51--72.

\bibitem{Arons79}
J.~{Arons} and E.~T. {Scharlemann}, {\it {Pair formation above pulsar polar
  caps - Structure of the low altitude acceleration zone}},  {\em ApJ} {\bf
  231} (Aug., 1979) 854--879.

\bibitem{Arons02}
J.~{Arons}, {\it {Theory of Pulsar Winds}},  in {\em Neutron Stars in Supernova
  Remnants} (P.~O. {Slane} and B.~M. {Gaensler}, eds.), vol.~271 of {\em
  Astronomical Society of the Pacific Conference Series}, p.~71, 2002.

\bibitem{Kirk09}
J.~G. {Kirk}, Y.~{Lyubarsky}, and J.~{Petri}, {\it {The Theory of Pulsar Winds
  and Nebulae}},  in {\em Astrophysics and Space Science Library} (W.~{Becker},
  ed.), vol.~357 of {\em Astrophysics and Space Science Library}, p.~421, 2009.
\newblock \href{http://xxx.lanl.gov/abs/astro-ph/}{{\tt astro-ph/}}.

\bibitem{LKP13}
M.~{Lemoine}, K.~{Kotera}, and J.~{P{\'e}tri}, {\it {On ultra-high energy
  cosmic ray acceleration at the termination shock of young pulsar winds}},
  {\em ArXiv e-prints: 1409.0159} (Aug., 2014)
  [\href{http://xxx.lanl.gov/abs/1409.0159}{{\tt arXiv:1409.0159}}].

\bibitem{2041-8205-795-1-L22}
A.~Y. Chen and A.~M. Beloborodov, {\it Electrodynamics of axisymmetric pulsar
  magnetosphere with electron-positron discharge: A numerical experiment},
  {\em The Astrophysical Journal Letters} {\bf 795} (2014), no.~1 L22.

\bibitem{Goldreich69}
P.~{Goldreich} and W.~H. {Julian}, {\it {Pulsar Electrodynamics}},  {\em ApJ}
  {\bf 157} (Aug., 1969) 869.

\bibitem{2013MNRAS.431L..48P}
O.~{Porth}, S.~S. {Komissarov}, and R.~{Keppens}, {\it {Solution to the sigma
  problem of pulsar wind nebulae}},  {\em MNRAS} {\bf 431} (Apr., 2013)
  L48--L52, [\href{http://xxx.lanl.gov/abs/1212.1382}{{\tt arXiv:1212.1382}}].

\bibitem{2014MNRAS.437..703M}
B.~D. {Metzger}, I.~{Vurm}, R.~{Hasco{\"e}t}, and A.~M. {Beloborodov}, {\it
  {Ionization break-out from millisecond pulsar wind nebulae: an X-ray probe of
  the origin of superluminous supernovae}},  {\em MNRAS} {\bf 437} (Jan., 2014)
  703--720, [\href{http://xxx.lanl.gov/abs/1307.8115}{{\tt arXiv:1307.8115}}].

\bibitem{Murase08}
K.~{Murase}, K.~{Ioka}, S.~{Nagataki}, and T.~{Nakamura}, {\it {High-energy
  cosmic-ray nuclei from high- and low-luminosity gamma-ray bursts and
  implications for multimessenger astronomy}},  {\em Phys. Rev. D} {\bf 78}
  (July, 2008) 023005--+, [\href{http://xxx.lanl.gov/abs/0801.2861}{{\tt
  arXiv:0801.2861}}].

\bibitem{Wang08}
X.-Y. {Wang}, S.~{Razzaque}, and P.~{M{\'e}sz{\'a}ros}, {\it {On the Origin and
  Survival of Ultra-High-Energy Cosmic-Ray Nuclei in Gamma-Ray Bursts and
  Hypernovae}},  {\em ApJ} {\bf 677} (Apr., 2008) 432--440,
  [\href{http://xxx.lanl.gov/abs/0711.2065}{{\tt arXiv:0711.2065}}].

\bibitem{2011MNRAS.415.2495M}
B.~D. {Metzger}, D.~{Giannios}, and S.~{Horiuchi}, {\it {Heavy nuclei
  synthesized in gamma-ray burst outflows as the source of ultrahigh energy
  cosmic rays}},  {\em MNRAS} {\bf 415} (Aug., 2011) 2495--2504,
  [\href{http://xxx.lanl.gov/abs/1101.4019}{{\tt arXiv:1101.4019}}].

\bibitem{Horiuchi12}
S.~{Horiuchi}, K.~{Murase}, K.~{Ioka}, and P.~{M{\'e}sz{\'a}ros}, {\it {The
  Survival of Nuclei in Jets Associated with Core-collapse Supernovae and
  Gamma-Ray Bursts}},  {\em ApJ} {\bf 753} (July, 2012) 69,
  [\href{http://xxx.lanl.gov/abs/1203.0296}{{\tt arXiv:1203.0296}}].

\bibitem{Usov92}
V.~V. {Usov}, {\it {Millisecond pulsars with extremely strong magnetic fields
  as a cosmological source of gamma-ray bursts}},  {\em Nature} {\bf 357}
  (June, 1992) 472--474.

\bibitem{Bonazzola96}
S.~{Bonazzola} and E.~{Gourgoulhon}, {\it {Gravitational waves from pulsars:
  emission by the magnetic-field-induced distortion.}},  {\em A\&A} {\bf 312}
  (Aug., 1996) 675--690, [\href{http://xxx.lanl.gov/abs/astro-ph/}{{\tt
  astro-ph/}}].

\bibitem{Ostriker69}
J.~P. {Ostriker} and J.~E. {Gunn}, {\it {On the Nature of Pulsars. I. Theory}},
   {\em ApJ} {\bf 157} (Sept., 1969) 1395--+.

\bibitem{KPO13}
K.~{Kotera}, E.~S. {Phinney}, and A.~V. {Olinto}, {\it {Signatures of pulsars
  in the light curves of newly formed supernova remnants}},  {\em MNRAS} {\bf
  432} (July, 2013) 3228--3236, [\href{http://xxx.lanl.gov/abs/1304.5326}{{\tt
  arXiv:1304.5326}}].

\bibitem{Chevalier05}
R.~A. {Chevalier}, {\it {Young Core-Collapse Supernova Remnants and Their
  Supernovae}},  {\em ApJ} {\bf 619} (Feb., 2005) 839--855,
  [\href{http://xxx.lanl.gov/abs/astro-ph/}{{\tt astro-ph/}}].

\bibitem{Rees74}
M.~J. {Rees} and J.~E. {Gunn}, {\it {The origin of the magnetic field and
  relativistic particles in the Crab Nebula}},  {\em MNRAS} {\bf 167} (Apr.,
  1974) 1--12.

\bibitem{Kennel84}
C.~F. {Kennel} and F.~V. {Coroniti}, {\it {Confinement of the Crab pulsar's
  wind by its supernova remnant}},  {\em ApJ} {\bf 283} (Aug., 1984) 694--709.

\bibitem{Kennel84b}
C.~F. {Kennel} and F.~V. {Coroniti}, {\it {Magnetohydrodynamic model of Crab
  nebula radiation}},  {\em ApJ} {\bf 283} (Aug., 1984) 710--730.

\bibitem{Atoyan96}
A.~M. {Atoyan} and F.~A. {Aharonian}, {\it {On the mechanisms of gamma
  radiation in the Crab Nebula}},  {\em MNRAS} {\bf 278} (Jan., 1996) 525--541.

\bibitem{DelZanna04}
L.~{Del Zanna}, E.~{Amato}, and N.~{Bucciantini}, {\it {Axially symmetric
  relativistic MHD simulations of Pulsar Wind Nebulae in Supernova Remnants. On
  the origin of torus and jet-like features}},  {\em A\&A} {\bf 421} (July,
  2004) 1063--1073, [\href{http://xxx.lanl.gov/abs/astro-ph/}{{\tt
  astro-ph/}}].

\bibitem{Komissarov04}
S.~{Komissarov} and Y.~{Lyubarsky}, {\it {MHD Simulations of Crab's Jet and
  Torus}},  {\em Ap \& SS} {\bf 293} (Sept., 2004) 107--113.

\bibitem{Murase:2007yt}
K.~Murase, {\it {High energy neutrino early afterglows gamma-ray bursts
  revisited}},  {\em Phys.Rev.} {\bf D76} (2007) 123001,
  [\href{http://xxx.lanl.gov/abs/0707.1140}{{\tt arXiv:0707.1140}}].

\bibitem{2015arXiv151205368L}
P.~D. {Lasky} and K.~{Glampedakis}, {\it {Observationally constraining
  gravitational wave emission from short gamma-ray burst remnants}},  {\em
  ArXiv e-prints} (Dec., 2015) [\href{http://xxx.lanl.gov/abs/1512.0536}{{\tt
  arXiv:1512.0536}}].

\bibitem{Matzner99}
C.~D. {Matzner} and C.~F. {McKee}, {\it {The Expulsion of Stellar Envelopes in
  Core-Collapse Supernovae}},  {\em ApJ} {\bf 510} (Jan., 1999) 379--403,
  [\href{http://xxx.lanl.gov/abs/astro-ph/}{{\tt astro-ph/}}].

\bibitem{Apel:2011mi}
W.~Apel et~al., {\it {Kneelike structure in the spectrum of the heavy component
  of cosmic rays observed with KASCADE-Grande}},  {\em Phys.Rev.Lett.} {\bf
  107} (2011) 171104, [\href{http://xxx.lanl.gov/abs/1107.5885}{{\tt
  arXiv:1107.5885}}].

\bibitem{Apel:2013ura}
W.~Apel, J.~Arteaga-Vel{\`a}zquez, K.~Bekk, M.~Bertaina, J.~Bl{\"u}mer, et~al.,
  {\it {Ankle-like Feature in the Energy Spectrum of Light Elements of Cosmic
  Rays Observed with KASCADE-Grande}},  {\em Phys.Rev.} {\bf D87} (2013)
  081101, [\href{http://xxx.lanl.gov/abs/1304.7114}{{\tt arXiv:1304.7114}}].

\bibitem{ThePierreAuger:2013eja}
{\bf Pierre Auger Collaboration} Collaboration, A.~Aab et~al., {\it {The Pierre
  Auger Observatory: Contributions to the 33rd International Cosmic Ray
  Conference (ICRC 2013)}},  \href{http://xxx.lanl.gov/abs/1307.5059}{{\tt
  arXiv:1307.5059}}.

\bibitem{Abbasi:2011zx}
R.~Abbasi et~al., {\it {The IceCube Neutrino Observatory II: All Sky Searches:
  Atmospheric, Diffuse and EHE}},
  \href{http://xxx.lanl.gov/abs/1111.2736}{{\tt arXiv:1111.2736}}.

\bibitem{2013ApJ...779..132A}
M.~G. {Aartsen}, R.~{Abbasi}, Y.~{Abdou}, M.~{Ackermann}, J.~{Adams}, J.~A.
  {Aguilar}, M.~{Ahlers}, D.~{Altmann}, J.~{Auffenberg}, X.~{Bai}, and et~al.,
  {\it {Search for Time-independent Neutrino Emission from Astrophysical
  Sources with 3 yr of IceCube Data}},  {\em ApJ} {\bf 779} (Dec., 2013) 132.

\bibitem{F15}
K.~{Fang}, {\it {High-energy neutrino signatures of newborn pulsars in the
  local universe}},  {\em JCAP} {\bf 6} (June, 2015) 4,
  [\href{http://xxx.lanl.gov/abs/1411.2174}{{\tt arXiv:1411.2174}}].

\bibitem{2006ApJ...651..142H}
A.~M. {Hopkins} and J.~F. {Beacom}, {\it {On the Normalization of the Cosmic
  Star Formation History}},  {\em ApJ} {\bf 651} (Nov., 2006) 142--154,
  [\href{http://xxx.lanl.gov/abs/astro-ph/0601463}{{\tt astro-ph/0601463}}].

\bibitem{L05}
M.~{Lemoine}, {\it {Extragalactic magnetic fields and the second knee in the
  cosmic-ray spectrum}},  {\em Phys. Rev. D} {\bf 71} (Apr., 2005) 083007,
  [\href{http://xxx.lanl.gov/abs/0411173}{{\tt 0411173}}].

\bibitem{KL08a}
K.~{Kotera} and M.~{Lemoine}, {\it {Inhomogeneous extragalactic magnetic fields
  and the second knee in the cosmic ray spectrum}},  {\em Phys. Rev. D} {\bf
  77} (Jan., 2008) 023005, [\href{http://xxx.lanl.gov/abs/0706.1891}{{\tt
  arXiv:0706.1891}}].

\bibitem{KL08b}
K.~{Kotera} and M.~{Lemoine}, {\it {Optical depth of the Universe to ultrahigh
  energy cosmic ray scattering in the magnetized large scale structure}},  {\em
  Phys. Rev. D} {\bf 77} (June, 2008) 123003,
  [\href{http://xxx.lanl.gov/abs/0801.1450}{{\tt arXiv:0801.1450}}].

\bibitem{2003ApJ...595..346Z}
B.~{Zhang}, Z.~G. {Dai}, P.~{M{\'e}sz{\'a}ros}, E.~{Waxman}, and A.~K.
  {Harding}, {\it {High-Energy Neutrinos from Magnetars}},  {\em ApJ} {\bf 595}
  (Sept., 2003) 346--351, [\href{http://xxx.lanl.gov/abs/astro-ph/0210382}{{\tt
  astro-ph/0210382}}].

\bibitem{Haensel99}
P.~{Haensel}, J.~P. {Lasota}, and J.~L. {Zdunik}, {\it On the minimum period of
  uniformly rotating neutron stars},  {\em Astronomy and Astrophysics} {\bf
  344} (Apr., 1999) 151--153.

\bibitem{2004ApJ...611..380T}
T.~A. {Thompson}, P.~{Chang}, and E.~{Quataert}, {\it {Magnetar Spin-Down,
  Hyperenergetic Supernovae, and Gamma-Ray Bursts}},  {\em ApJ} {\bf 611}
  (Aug., 2004) 380--393, [\href{http://xxx.lanl.gov/abs/astro-ph/0401555}{{\tt
  astro-ph/0401555}}].

\bibitem{Komissarov07}
S.~S. {Komissarov} and M.~V. {Barkov}, {\it {Magnetar-energized supernova
  explosions and gamma-ray burst jets}},  {\em MNRAS} {\bf 382} (Dec., 2007)
  1029--1040, [\href{http://xxx.lanl.gov/abs/0707.0264}{{\tt
  arXiv:0707.0264}}].

\bibitem{Bucciantini07}
N.~{Bucciantini}, E.~{Quataert}, J.~{Arons}, B.~D. {Metzger}, and T.~A.
  {Thompson}, {\it {Magnetar-driven bubbles and the origin of collimated
  outflows in gamma-ray bursts}},  {\em MNRAS} {\bf 380} (Oct., 2007)
  1541--1553, [\href{http://xxx.lanl.gov/abs/0705.1742}{{\tt
  arXiv:0705.1742}}].

\bibitem{Bucciantini08}
N.~{Bucciantini}, E.~{Quataert}, J.~{Arons}, B.~D. {Metzger}, and T.~A.
  {Thompson}, {\it {Relativistic jets and long-duration gamma-ray bursts from
  the birth of magnetars}},  {\em MNRAS} {\bf 383} (Jan., 2008) L25--L29,
  [\href{http://xxx.lanl.gov/abs/0707.2100}{{\tt arXiv:0707.2100}}].

\bibitem{Bucciantini09}
N.~{Bucciantini}, E.~{Quataert}, B.~D. {Metzger}, T.~A. {Thompson}, J.~{Arons},
  and L.~{Del Zanna}, {\it {Magnetized relativistic jets and long-duration GRBs
  from magnetar spin-down during core-collapse supernovae}},  {\em MNRAS} {\bf
  396} (July, 2009) 2038--2050, [\href{http://xxx.lanl.gov/abs/0901.3801}{{\tt
  arXiv:0901.3801}}].

\bibitem{Metzger11}
B.~D. {Metzger}, D.~{Giannios}, and S.~{Horiuchi}, {\it {Heavy nuclei
  synthesized in gamma-ray burst outflows as the source of ultrahigh energy
  cosmic rays}},  {\em MNRAS} {\bf 415} (Aug., 2011) 2495--2504,
  [\href{http://xxx.lanl.gov/abs/1101.4019}{{\tt arXiv:1101.4019}}].

\bibitem{Begelman92}
M.~C. {Begelman} and Z.-Y. {Li}, {\it {An axisymmetric magnetohydrodynamic
  model for the Crab pulsar wind bubble}},  {\em ApJ} {\bf 397} (Sept., 1992)
  187--195.

\bibitem{Murase:2010gj}
K.~Murase and J.~F. Beacom, {\it {Neutrino Background Flux from Sources of
  Ultrahigh-Energy Cosmic-Ray Nuclei}},  {\em Phys.Rev.} {\bf D81} (2010)
  123001, [\href{http://xxx.lanl.gov/abs/1003.4959}{{\tt arXiv:1003.4959}}].

\bibitem{2010ApJ...717..245K}
D.~{Kasen} and L.~{Bildsten}, {\it {Supernova Light Curves Powered by Young
  Magnetars}},  {\em ApJ} {\bf 717} (July, 2010) 245--249,
  [\href{http://xxx.lanl.gov/abs/0911.0680}{{\tt arXiv:0911.0680}}].

\bibitem{2011Natur.474..487Q}
R.~M. {Quimby}, S.~R. {Kulkarni}, M.~M. {Kasliwal}, A.~{Gal-Yam}, I.~{Arcavi},
  M.~{Sullivan}, P.~{Nugent}, R.~{Thomas}, D.~A. {Howell}, E.~{Nakar},
  L.~{Bildsten}, C.~{Theissen}, N.~M. {Law}, R.~{Dekany}, G.~{Rahmer},
  D.~{Hale}, R.~{Smith}, E.~O. {Ofek}, J.~{Zolkower}, V.~{Velur}, R.~{Walters},
  J.~{Henning}, K.~{Bui}, D.~{McKenna}, D.~{Poznanski}, S.~B. {Cenko}, and
  D.~{Levitan}, {\it {Hydrogen-poor superluminous stellar explosions}},  {\em
  Nature} {\bf 474} (June, 2011) 487--489,
  [\href{http://xxx.lanl.gov/abs/0910.0059}{{\tt arXiv:0910.0059}}].

\bibitem{Faucher06}
C.-A. {Faucher-Gigu{\`e}re} and V.~M. {Kaspi}, {\it {Birth and Evolution of
  Isolated Radio Pulsars}},  {\em ApJ} {\bf 643} (May, 2006) 332--355.

\end{thebibliography}\endgroup

\vfill\eject
%\end\bye
\end{document}